\documentclass{article}

\usepackage{microtype}
\usepackage{graphicx}
\usepackage{subcaption}
\usepackage{booktabs} %

\usepackage{hyperref}

\usepackage[accepted]{icml2026}

\usepackage{amsmath}
\usepackage{amssymb}
\usepackage{mathtools}
\usepackage{amsthm}

\usepackage[capitalize,noabbrev]{cleveref}

\theoremstyle{plain}

\theoremstyle{definition}

\theoremstyle{remark}

\usepackage[textsize=tiny]{todonotes}

\usepackage{microtype}
\usepackage{booktabs}
\usepackage{amsmath,amssymb,amsfonts}
\usepackage{nicefrac}
\usepackage[table]{xcolor}
\usepackage{wrapfig}
\usepackage{outlines}
\usepackage{enumitem}
\setlist{leftmargin=*}
\setitemize{leftmargin=*}

\definecolor{ETHBlue}{RGB}{33,92,175}
\definecolor{ETHGreen}{RGB}{98,115,19}
\definecolor{ETHPurpleDark}{RGB}{140,10,89}
\definecolor{ETHPurple}{RGB}{163,7,116}
\definecolor{ETHPurpleLight}{RGB}{220, 158, 201}
\definecolor{ETHGray}{RGB}{111,111,111}
\definecolor{ETHRed}{RGB}{183,53,45}
\definecolor{ETHPetrol}{RGB}{0,120,148}
\definecolor{ETHBronze}{RGB}{142,103,19}

\usepackage{etoolbox}
\newtoggle{color-macro}
\settoggle{color-macro}{false}

\iftoggle{color-macro}{
  \colorlet{MacroColor}{ETHPetrol}
}{
  \colorlet{MacroColor}{black}
}

\usepackage{amsthm}

\usepackage{tcolorbox}
\usepackage{tikz}
\usepackage{pgfplots}
\pgfplotsset{compat=1.18}
\usepgfplotslibrary{fillbetween}
\usepackage{scalefnt}
\usepackage{anyfontsize}
\usepackage{multirow}
\usepackage{changepage}
\usepackage{multicol}
\usepackage{setspace}
\usepackage[version=4]{mhchem}

\usepackage{caption}
\usepackage[font=small,labelfont=bf]{subcaption}

\usepackage{url}
\usepackage{hyperref}

\usepackage{cleveref}
\crefname{section}{\S}{\S\S}
\Crefname{section}{\S}{\S\S}
\crefname{table}{Tab.}{Tabs.}
\crefname{figure}{Fig.}{Figs.}
\crefname{algorithm}{Alg.}{}
\crefname{appendix}{App.}{Apps.}
\crefname{lemma}{Lemma}{}
\Crefname{theorem}{Theorem}{}
\crefname{proposition}{Proposition}{}
\crefname{hypothesis}{Hypothesis}{}
\crefname{deduction}{Deduction}{}
\crefname{intuition}{\textbf{Intuition}}{\textbf{Intuitions}}
\crefname{observation}{\textbf{Observation}}{\textbf{Observations}}
\crefname{finding}{\textbf{Finding}}{\textbf{Findings}}
\crefname{cor}{Corollary}{}
\crefname{align}{}{}
\crefname{equation}{}{}

\usepackage{xurl}

\icmltitlerunning{Unveiling the Visual Counting Bottleneck in Vision-Language Models}

\begin{document}

\twocolumn[
  \icmltitle{Unveiling the Visual Counting Bottleneck in Vision-Language Models}

  \icmlsetsymbol{equal}{*}

  \begin{icmlauthorlist}
    \icmlauthor{Xingzhou Pang}{equal,ethz}
    \icmlauthor{Yifan Hou}{equal,ethz}
    \icmlauthor{Junling Wang}{ethz}
    \icmlauthor{Mrinmaya Sachan}{ethz}
    \\
    $\{$
    \texttt{\href{mailto:xipang@ethz.ch}{xingzhou.pang},}
    \texttt{\href{mailto:yifan.hou@inf.ethz.ch}{yifan.hou},}
    \texttt{\href{mailto:junling.wang@ai.ethz.ch}{junling.wang},}
    \texttt{\href{mailto:mrinmaya.sachan@inf.ethz.ch}{mrinmaya.sachan}}
    $\}$\texttt{@inf.ethz.ch}
    \\
    {%
    \setlength{\fboxsep}{2.5pt}%
    \setlength{\fboxrule}{2.5pt}%
    \fcolorbox{white}{white}{\includegraphics[width=.15\linewidth]{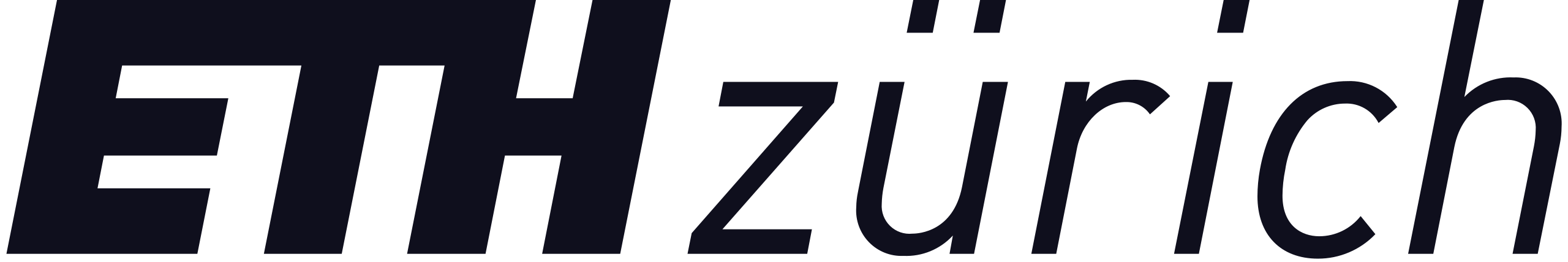}}
    }
  \end{icmlauthorlist}

  \icmlaffiliation{ethz}{Department of Computer Science, ETH Zürich}

  \icmlcorrespondingauthor{Yifan Hou}{yifan.hou@inf.ethz.ch}
  \icmlcorrespondingauthor{Mrinmaya Sachan}{mrinmaya.sachan@inf.ethz.ch}

  \icmlkeywords{Machine Learning, ICML}

  \vskip 0.3in
]

\printAffiliationsAndNotice{\icmlEqualContribution}

\begin{abstract}
    While Large Vision-Language Models (VLMs) excel at interpolation, they suffer catastrophic failures in systematic generalization, most notably in visual counting.
    In this work, we investigate this extrapolation bottleneck by deconstructing visual counting into three cognitive stages: \textit{visual individuation}, \textit{magnitude awareness}, and \textit{symbolic mapping}.
    Using synthetic Go boards and linear probes, we demonstrate that visual backbones maintain robust, linearly separable representations of quantity well into the extrapolation regime, ruling out perceptual failure.
    Furthermore, models retain latent magnitude awareness, successfully performing comparative reasoning on quantities they fail to enumerate.
    We pinpoint the collapse to the \textit{symbolic mapping} stage, where the model fails to project valid visual magnitudes onto symbolic tokens.   
    Our findings support a \textit{fractured magnitude hypothesis}: VLMs fail to acquire a \textit{universal number space}, instead learning disjoint, modality-specific statistical manifolds that prevent cross-modal grounding for unseen quantities.
    Validated on the state-of-the-art foundation model, our results suggest that bridging this gap requires inductive priors enforcing unified representations, as data scaling alone is insufficient.\footnote{Our code is publicly available here: \url{https://github.com/Russellpang/semproj}.}
\end{abstract}

\section{Introduction}
Systematic generalization, the ability to learn a rule from finite examples and apply it to inputs outside the training distribution, remains the central chasm between biological and artificial intelligence~\citep{fodor1988connectionism,marcus2003algebraic,systemgeneralization_lake_icml2018,composedecompose_hupkes_jair2020}.
On the one hand, while Large Vision-Language Models (VLMs) have demonstrated impressive proficiency in describing visual scenes and solving visual reasoning~\citep{gpt_openai_arxiv2023,gemini_google_arxiv2023,gemma3_google_arxiv2025,qwen3vl_bai_arxiv2025}, others claim they are mere statistical interpolators, often excelling within the support of their training distribution but are error-prone when required to extrapolate~\citep{probevlm_thrush_cvpr2022,vlmbow_yuksekgonul_iclr2023,clipcount_paiss_iccv2023}.
This limitation is most clearly evident in the task of \textit{visual counting}.

Counting serves as a canonical testbed for reasoning as it isolates the problem of extrapolation in its purest form.
Grounded in a simple recursive algorithm ($n \rightarrow n+1$), counting allows humans to enumerate arbitrary quantities zero-shot once the principles of cardinality are acquired~\citep{dedekind1965sollen,carey2000origin,dehaene2011number,piantadosi2012bootstrapping}.
In contrast, neural models treat counting as a pattern-matching problem, degrading catastrophically when object quantities exceed those observed during training~\citep{nlpnum_wallace_emnlp2019,climbnlu_bender_acl2020,traintestlen_press_iclr2022,explorelen_anil_nips2022}.
This failure raises a fundamental diagnostic question: \textit{Does the failure to count stem from an inability to perceive distinct objects, %
an inability to comprehend quantity, %
or an inability to map that quantity to a label?} %

\begin{figure}[!t]
    \centering
    \includegraphics[width=\linewidth]{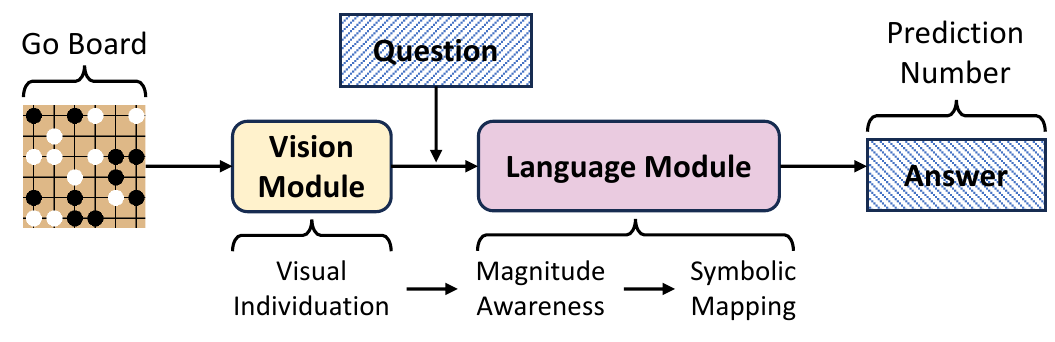}
    \caption{\textbf{Deconstructing visual counting.} We analyze the counting bottleneck in VLMs via 3 cognitive stages: (1) visual individuation, (2) magnitude awareness, and (3) symbolic mapping.}    
    \label{fig:three_stages}
\end{figure}

We investigate the bottleneck of visual counting in VLMs by deconstructing it into three cognitive stages (\cref{fig:three_stages}): \textit{visual individuation} (recognition), \textit{magnitude awareness} (numerosity), and \textit{symbolic mapping} (articulation).
To rigorously isolate architectural biases from the noise inherent in real-world datasets, we construct a synthetic laboratory using Go game boards (\cref{sec:framework:syn}).
Crucially, we employ a decoupled training curriculum similar to VLM training: the model is trained to visually count only up to $N=49$, while its language decoder is pretrained to count textually up to $N=99$.
This creates a specific \textit{visual extrapolation} regime ($50-99$), where the model possesses the linguistic labels but with a lack of visual-textual pairings, and a \textit{full extrapolation} scenario ($100-120$) where both the visual densities and the textual quantities are unseen.
We further validate our findings on a state-of-the-art VLM (Qwen3-VL) to confirm that the observed failure mechanisms persist in real-world architectures regardless of pretraining scale (\cref{sec:framework:real}).

To measure the model's latent counting capability, we operationalize the \textit{hidden number} via the linear separability of its internal representations. 
We posit that if feature vectors for distinct quantities remain linearly distinguishable, the model inherently possesses the capacity for \textit{visual individuation}. 
Our investigation reveals a counter-intuitive phenomenon: the bottleneck lies neither in visual recognition nor in abstract numerosity. 
Visual backbones maintain robust, linearly separable representations of quantity well beyond the training distribution, confirming that the model effectively ``sees'' the objects (\cref{sec:syn:stage1}). 
Furthermore, models retain latent \textit{magnitude awareness}, successfully performing comparative reasoning on quantities far outside the training set (e.g., accurately determining if a visual set of 110 items corresponds to a textual list of 110 characters) even when explicit enumeration fails (\cref{sec:syn:stage2}).
Instead, the generalization collapse is pinpointed strictly to the final \textit{symbolic mapping} stage (\cref{sec:syn:stage3}). 
Validation on foundation models confirms this mechanism: while real-world architectures exhibit minor noise in perception and magnitude processing, the catastrophic failure is consistently driven by the structural disconnect between the preserved latent magnitude and the generated symbolic token (\cref{sec:real}).

We interpret these findings through the lens of a \textit{fractured magnitude hypothesis}.
We posit that current VLM architectures fail to acquire a \textit{universal number space} that bridges modalities; instead, they learn disjoint, modality-specific statistical manifolds for visual and textual magnitudes.
Consequently, while the language module may possess a robust representation of a number (e.g., 75) and the vision module possesses a precise representation of the corresponding objects, the model fails to map the latter to the former if that specific cross-modal pairing is not observed during training.
These findings suggest that solving the extrapolation gap in VLMs is not merely a function of scaling data, but requires architectural or training priors that enforce unified magnitude representations to enable true cross-modal grounding.

\section{Experimental Framework} 
\label{sec:framework}
To rigorously investigate the visual counting bottleneck, we adopt a two-fold experimental design.
First, we construct a \textit{synthetic laboratory} using a custom VLM trained from scratch, allowing us to strictly control the training distribution and decouple visual exposure from textual priors (\cref{sec:framework:syn}).
Second, we perform a \textit{real-world validation} using a state-of-the-art pretrained VLM to confirm that our findings hold in foundation models trained at scale (\cref{sec:framework:real}).

\subsection{Study 1: The Synthetic Laboratory} 
\label{sec:framework:syn}

\paragraph{Architecture.} 
For initial analysis, we train a lightweight ``Toy VLM'' composed of standard architectural primitives: a vision Transformer encoder (ViT-Base configuration but only with 2 layers) for visual perception~\citep{vit_dosovitskiy_iclr2021}, connected to a causal Transformer decoder (GPT-2 style, 2 layers) for text generation~\citep{radford2019language}.
This mirrors the architectural bias of modern VLMs (e.g., Qwen-VL~\citep{qwen3vl_bai_arxiv2025}, Gemma~\citep{gemma3_google_arxiv2025}) while remaining computationally tractable for full retraining.

\paragraph{The Go Board Dataset.}
We utilize standard $19 \times 19$ Go game boards as a controlled environment for testing and evaluating our model. 
Unlike natural images confounded by occlusion or semantic ambiguity, this setup offers precise, deterministic control over object density in the generated images.
We generate board configurations where target objects (black stones) are placed at random, non-overlapping coordinates.
To preclude solutions based on global pixel statistics (e.g., total foreground area), we introduce distractor objects (white stones) with cardinalities varying by up to $\pm 30$ relative to the target; this forces the model to perform specific feature attention rather than global density estimation.
Crucially, we calibrate the stone sizes ($14 \times 14$ pixels) to match the VLM's input patch size, ensuring a theoretically clean mapping between distinct objects and token representations.

\paragraph{Decoupled Training Curriculum.}
To test whether linguistic priors automatically ground unseen visual inputs, we emulate VLM pretraining dynamics via a decoupled curriculum that creates a deliberate distribution shift between modalities.
In \textit{Phase 1 (Language Pretraining)}, we equip the decoder with symbolic counting capabilities by training exclusively on text sequences (length $361$) containing up to $N=99$ target characters (see \cref{fig:prompt_template:syn:text} in \cref{app:sup:settings:prompt} for details).
This ensures the model possesses the requisite algorithmic priors and output tokens independent of vision.

In \textit{Phase 2 (Vision Alignment)}, we finetune the full VLM on visual counting but strictly limit the dataset to $N \in [0, 49]$ (see \cref{fig:prompt_template:syn:vision} in \cref{app:sup:settings:prompt} for details).
Crucially, to prevent catastrophic forgetting of the linguistic priors, we co-train the model with the textual counting dataset ($N \in [0, 99]$) during this phase.
This truncation creates a critical visual extrapolation regime ($50-99$): a zone where the model possesses the labels (from Phase 1) but has never observed their corresponding visual densities.
To preserve the textual priors during alignment, we apply a reduced learning rate to the decoder. Full hyperparameters are provided in \cref{app:sup:settings:training}.

\paragraph{Evaluation Protocol.}
We evaluate the VLM on three sets:
(1) \textit{In-Distribution (ID, $0-49$):} Interpolation within the visual training support.
(2) \textit{Visual Extrapolation (VE, $50-99$):} The critical regime where visual density is novel but labels are known. Failure here indicates a failure of cross-modal grounding.
(3) \textit{Full Extrapolation (FE, $100-120$):} A regime where both visual density and text labels are unseen.
For prediction, we employ standard greedy decoding with an End-of-Sequence (EOS) token, treating counting as an open-ended generation task rather than classification.

\subsection{Study 2: Validation on a State-of-the-Art VLM} 
\label{sec:framework:real}

\paragraph{Model Selection.}
We extend our analysis to Qwen3-VL-32B-Instruct~\citep{qwen3vl_bai_arxiv2025}, a state-of-the-art open-weights VLM that exhibits strong visual counting capabilities. 
We focus exclusively on Qwen3-VL because meaningful extrapolation analysis requires a baseline competence; other open-source models failed to achieve sufficient accuracy even in the low-number regime ($N < 10$) on our diagnostic task. 
Analyzing this architecture allows us to verify if the ``fractured magnitude'' phenomenon persists despite trillion-token scale pretraining and instruction tuning.

\paragraph{Dataset Configuration.}
We adapt the synthetic Go board generator to a $6\times 6$ grid format.
Correspondingly, the target object number ranges from 0 to 20, with distractors (white stones) up to $\pm 5$ relative to the target.
This resolution is empirically calibrated to the effective capacity of the pretrained model, ensuring the task complexity lies at the boundary of its capabilities. 
To facilitate precise probing, we align the object dimensions with the model's patch processing: since Qwen3-VL merges $2\times2$ sets of $16\times16$ patches into a single token, we scale each stone to $32\times32$ pixels to strictly align object boundaries with token inputs.

\paragraph{Prompt Design.}
We evaluate the model in a zero-shot setting using Chain-of-Thought (CoT) prompting to elicit maximal reasoning performance. 
Full prompt templates and generated response examples are detailed in \cref{app:sup:settings:prompt}.

\subsection{Diagnostic Tool: Hidden Number Probing}
\label{sec:framework:probe}

To diagnose \textit{why} a model fails (blindness vs. decoding failure), we employ linear probing to extract the \textit{hidden number} ($N_H$): the quantity information physically present in the visual token representations.

Let $\mathbf{Z} = \{z_1, z_2, \dots, z_L\}$ denote the sequence of patch output ($d$-dimension) embeddings from the visual encoder $\mathcal{M}_V$, e.g., $L=19\times19=361$ corresponds to the number of visual tokens.
We train a binary linear probe $f_{\text{probe}}: \mathbb{R}^d \rightarrow \{0, 1\}$ to detect the presence of a black stone at the grid position corresponding to token $z_i$.
Crucially, $f_{\text{probe}}$ is trained exclusively on the ID dataset ($N_G \le 49$) to test generalization.
The \textit{hidden number} ($N_H$) is obtained by aggregating the probe predictions across the full sequence:
\begin{equation}
    N_H = \sum_{i=1}^{L} f_{\text{probe}}(z_i).
\end{equation}
While defined here for the visual encoder output, this probing methodology can be applied to the hidden states of any layer in the network.
By comparing the internal percept ($N_H$), the generated symbol ($N_P$), and the ground truth ($N_G$), we isolate the failure mechanism via two metrics:

\begin{itemize}[leftmargin=*]
    \item Vision Gap (i.e., recognition error), formally $ |N_H - N_G|$, which measures the errors brought by the vision module. If $N_H \approx N_G$, it means there is no error in vision module.
    \item Language Gap (i.e., numerosity and articulation error), formally $|N_H - N_P|$, which measures the errors brought by the language module. If $N_H \approx N_P$, it means there is no such error in the language module.
\end{itemize}

\section{Synthetic Extrapolation Laboratory}
\label{sec:syn}
In this section, we deconstruct the mechanics of visual counting failure using our controlled synthetic environment. 
We first quantify the extent of the generalization gap (\cref{sec:syn:acc}). 
We then systematically investigate the three hypothesized bottlenecks: \textit{visual individuation} (\cref{sec:syn:stage1}), \textit{magnitude awareness} (\cref{sec:syn:stage2}), and \textit{symbolic mapping} (\cref{sec:syn:stage3}).

\subsection{The Baseline Paradox}
\label{sec:syn:acc}

We aim to determine whether a VLM can zero-shot transfer its algorithmic counting knowledge from language to vision. 
Specifically, if a model has acquired the recursive rule of counting in the text domain (up to $N=99$) but has strictly limited visual exposure (up to $N=49$), can it bridge this modality gap to enumerate 50 objects? 
This establishes the baseline behavior for our \textit{Visual Extrapolation (VE)} regime.

\paragraph{Experiment Design.}
We evaluate the synthetic VLM on two parallel tasks across the full range of $N \in [0, 120]$ to isolate modality-specific capabilities.
For \textit{textual counting}, the model is prompted to count specific characters within a text sequence, testing its counting ability in the language domain.
For \textit{visual counting}, the model must predict the total number of black stones on a provided Go board image.
We report (Exact Match) accuracy across three defined regimes: In-Distribution (ID, $0-49$), Visual Extrapolation (VE, $50-99$), and Full Extrapolation (FE, $100-120$).

\begin{figure}[!t]
    \centering
    \scalefont{0.8}
    \begin{tikzpicture}
    \begin{axis}[
        width=8cm, height=5cm,
        xlabel={Number of Objects},
        ylabel={Counting Accuracy (\%)},
        xmin=0, xmax=120,
        ymin=-5, ymax=120,
        xtick={0, 20, 40, 60, 80, 100, 120},
        xticklabels={0, 20, 40, 60, 80, 100, 120},
        xtick pos=bottom,
        tick align=outside,
        xticklabel style={font=\scriptsize},
        ylabel near ticks,
        xlabel near ticks,
        legend style={at={(0.86,1.4)}, legend columns=2, anchor=north east, nodes={scale=0.9, transform shape},fill=white}, 
        ymajorgrids=true,
        grid style=dashed,
        clip=false
    ]

    \addplot+[
        solid,
        line width=0.4mm,
        color=ETHRed, 
        mark=none,
    ] coordinates {
(0, 100.0)
(1, 100.0)
(2, 100.0)
(3, 100.0)
(4, 100.0)
(5, 100.0)
(6, 100.0)
(7, 100.0)
(8, 100.0)
(9, 99.8046875)
(10, 99.31640625)
(11, 100.0)
(12, 100.0)
(13, 100.0)
(14, 100.0)
(15, 100.0)
(16, 100.0)
(17, 100.0)
(18, 100.0)
(19, 99.8046875)
(20, 100.0)
(21, 100.0)
(22, 100.0)
(23, 100.0)
(24, 100.0)
(25, 100.0)
(26, 100.0)
(27, 100.0)
(28, 100.0)
(29, 99.609375)
(30, 98.6328125)
(31, 100.0)
(32, 100.0)
(33, 100.0)
(34, 100.0)
(35, 100.0)
(36, 100.0)
(37, 100.0)
(38, 99.90234375)
(39, 99.31640625)
(40, 99.90234375)
(41, 100.0)
(42, 100.0)
(43, 100.0)
(44, 100.0)
(45, 100.0)
(46, 100.0)
(47, 100.0)
(48, 99.90234375)
(49, 99.90234375)
(50, 0.0)
(51, 0.0)
(52, 0.0)
(53, 0.0)
(54, 0.0)
(55, 0.0)
(56, 0.0)
(57, 0.0)
(58, 0.0)
(59, 0.0)
(60, 0.0)
(61, 0.0)
(62, 0.0)
(63, 0.0)
(64, 0.0)
(65, 0.0)
(66, 0.0)
(67, 0.0)
(68, 0.0)
(69, 0.0)
(70, 0.0)
(71, 0.0)
(72, 0.0)
(73, 0.0)
(74, 0.0)
(75, 0.0)
(76, 0.0)
(77, 0.0)
(78, 0.0)
(79, 0.0)
(80, 0.0)
(81, 0.0)
(82, 0.0)
(83, 0.0)
(84, 0.0)
(85, 0.0)
(86, 0.0)
(87, 0.0)
(88, 0.0)
(89, 0.0)
(90, 0.0)
(91, 0.0)
(92, 0.0)
(93, 0.0)
(94, 0.0)
(95, 0.0)
(96, 0.0)
(97, 0.0)
(98, 0.0)
(99, 0.0)
(100, 0.0)
(101, 0.0)
(102, 0.0)
(103, 0.0)
(104, 0.0)
(105, 0.0)
(106, 0.0)
(107, 0.0)
(108, 0.0)
(109, 0.0)
(110, 0.0)
(111, 0.0)
(112, 0.0)
(113, 0.0)
(114, 0.0)
(115, 0.0)
(116, 0.0)
(117, 0.0)
(118, 0.0)
(119, 0.0)
(120, 0.0)
    };
    \addlegendentry{Vision Counting}

    \addplot+[
        draw=none,          %
        fill=ETHBronze,     %
        fill opacity=0.4,   %
        mark=none,
        area legend
    ] coordinates {
(0, 100.0)
(1, 100.0)
(2, 100.0)
(3, 100.0)
(4, 100.0)
(5, 100.0)
(6, 100.0)
(7, 99.90234375)
(8, 100.0)
(9, 100.0)
(10, 100.0)
(11, 100.0)
(12, 100.0)
(13, 100.0)
(14, 100.0)
(15, 100.0)
(16, 100.0)
(17, 100.0)
(18, 100.0)
(19, 99.90234375)
(20, 100.0)
(21, 100.0)
(22, 100.0)
(23, 100.0)
(24, 100.0)
(25, 100.0)
(26, 100.0)
(27, 100.0)
(28, 99.90234375)
(29, 100.0)
(30, 100.0)
(31, 100.0)
(32, 100.0)
(33, 100.0)
(34, 100.0)
(35, 100.0)
(36, 100.0)
(37, 100.0)
(38, 100.0)
(39, 100.0)
(40, 100.0)
(41, 100.0)
(42, 100.0)
(43, 100.0)
(44, 100.0)
(45, 100.0)
(46, 100.0)
(47, 100.0)
(48, 100.0)
(49, 99.8046875)
(50, 100.0)
(51, 100.0)
(52, 100.0)
(53, 100.0)
(54, 100.0)
(55, 100.0)
(56, 100.0)
(57, 100.0)
(58, 100.0)
(59, 99.51171875)
(60, 100.0)
(61, 100.0)
(62, 100.0)
(63, 100.0)
(64, 100.0)
(65, 100.0)
(66, 100.0)
(67, 100.0)
(68, 100.0)
(69, 99.609375)
(70, 99.609375)
(71, 100.0)
(72, 100.0)
(73, 100.0)
(74, 100.0)
(75, 100.0)
(76, 100.0)
(77, 100.0)
(78, 100.0)
(79, 98.73046875)
(80, 99.609375)
(81, 100.0)
(82, 100.0)
(83, 100.0)
(84, 100.0)
(85, 100.0)
(86, 100.0)
(87, 100.0)
(88, 100.0)
(89, 99.8046875)
(90, 98.73046875)
(91, 100.0)
(92, 100.0)
(93, 100.0)
(94, 100.0)
(95, 100.0)
(96, 100.0)
(97, 100.0)
(98, 99.8046875)
(99, 100.0)
(100, 0.0)
(101, 0.0)
(102, 0.0)
(103, 0.0)
(104, 0.0)
(105, 0.0)
(106, 0.0)
(107, 0.0)
(108, 0.0)
(109, 0.0)
(110, 0.0)
(111, 0.0)
(112, 0.0)
(113, 0.0)
(114, 0.0)
(115, 0.0)
(116, 0.0)
(117, 0.0)
(118, 0.0)
(119, 0.0)
(120, 0.0)
    }\closedcycle;
    \addlegendentry{Text Counting}

    \draw [decorate, decoration={brace, amplitude=5pt, raise=2pt}, color=ETHBlue, line width=1pt]
    (axis cs:0.5, 120) -- (axis cs:48.5, 120) 
    node [midway, above=8pt, color=ETHBlue, font=\scriptsize\bfseries] {ID};

    \draw [decorate, decoration={brace, amplitude=5pt, raise=2pt}, color=ETHGreen, line width=1pt]
    (axis cs:50.5, 120) -- (axis cs:98.5, 120) 
    node [midway, above=8pt, color=ETHGreen, font=\scriptsize\bfseries] {VE};

    \draw [decorate, decoration={brace, amplitude=5pt, raise=2pt}, color=ETHPurple, line width=1pt]
    (axis cs:100.5, 120) -- (axis cs:119.5, 120) 
    node [midway, above=8pt, color=ETHPurple, font=\scriptsize\bfseries] {FE};

    \draw [dashed, gray, opacity=0.5] (axis cs:49, -5) -- (axis cs:49, 120);
    \draw [dashed, gray, opacity=0.5] (axis cs:99, -5) -- (axis cs:99, 120);

    \end{axis}
    \end{tikzpicture}
    \caption{\textbf{The Baseline Paradox: Linguistic competence does not ensure visual ability.} 
    While textual counting (bronze area) generalizes perfectly to $N=99$, visual counting (red line) collapses to near-zero immediately outside the training distribution ($N>49$). 
    This stark dissociation in the Visual Extrapolation (VE) regime demonstrates that the model fails to apply its known symbolic labels to unseen visual quantities.}
    \label{fig:syn:acc}
\end{figure}
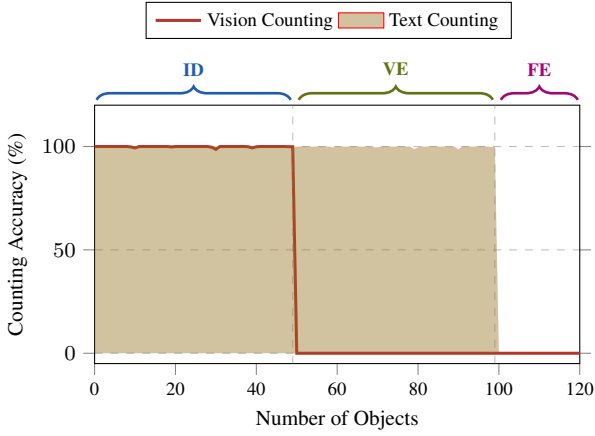
\paragraph{Experiment Results.}
As illustrated in \cref{fig:syn:acc}, the results reveal a stark modular dissociation.
The model achieves perfect accuracy ($100\%$) on the textual counting task throughout both the ID and VE regimes, confirming that the language decoder has successfully learned the recursive successor function up to $N=99$.
In contrast, visual counting performance matches this perfection only within the visual training support (ID).
Crucially, the moment the quantity exceeds the training boundary ($N > 49$), visual accuracy suffers a catastrophic collapse to $0\%$. 
Meanwhile, textual counting remains robust up to $N=99$ but abruptly collapses in the FE regime ($N \ge 100$), confirming that the model has interpolated within its text support rather than learning a length-agnostic recursive rule.

\paragraph{Takeaway.} 
These results establish the \textit{baseline paradox}: the model possesses the algorithmic capacity to count (proven by textual performance) yet appears effectively ``blind'' in the visual domain. 
The failure to extrapolate suggests a breakdown in the cross-modal processing chain. 
To pinpoint the bottleneck, we formulate three competing hypotheses regarding the locus of failure:
\begin{itemize}[leftmargin=*]
    \item \textbf{Hypothesis A (stage 1):} The bottleneck lies in \textit{visual individuation}. The visual encoder fails to resolve distinct object features in dense, unseen configurations, preventing the quantity signal from entering the system.
    \item \textbf{Hypothesis B (stage 2):} The bottleneck lies in \textit{magnitude awareness}. While the encoder captures the objects, the abstract quantity signal dissipates or degrades as it propagates through the deep language decoder layers.
    \item \textbf{Hypothesis C (stage 3):} The bottleneck lies in \textit{symbolic mapping}. The model retains the magnitude in its latent states but lacks the learned projection to map the visual representation to the corresponding symbolic token.
\end{itemize}

\subsection{Hypothesis A: Object Individuation}
\label{sec:syn:stage1}

The first logical suspect for the ``baseline paradox'' is \textit{perceptual crowding}: the inability of the visual encoder to distinguish individual objects in dense scenes that exceed training examples. 
If the model cannot ``see'' 75 objects, it certainly cannot count them.
To determine if the synthetic VLM is simply \textit{blind} to higher quantities, we bypass the language decoder and probe the information content of the visual output embeddings directly.

\paragraph{Experiment Design.}
We employ the \textit{hidden number probing} technique defined in \cref{sec:framework:probe}.
We train a linear probe ($f_{\text{probe}}$) on the frozen output of the visual encoder to predict object density.
Crucially, this probe is trained only on the ID data ($N \le 49$) to prevent any information leakage from the extrapolation regimes.
We then evaluate this probe across all three sets (ID, VE, FE) to extract the \textit{hidden number} ($N_H$) and compute our two diagnostic metrics:
(1) The \textit{Vision Gap} ($|N_H - N_G|$): measuring perceptual error.
(2) The \textit{Language Gap} ($|N_H - N_P|$): measuring the alignment error between the encoder's percept and the final token.

\begin{figure}[!t]
    \centering
    \scalefont{0.8}
    \begin{tikzpicture}
    \begin{axis}[
        width=8cm, height=5cm,
        xlabel={Number of Objects},
        ylabel={Number Gap},
        xmin=0, xmax=120,
        ymin=-2, ymax=10,
        xtick={0, 20, 40, 60, 80, 100, 120},
        xticklabels={0, 20, 40, 60, 80, 100, 120},
        ytick={0, 2, 4, 6, 8, 10},
        yticklabels={0, 2, 4, 6, 8, $\ge$10},
        xtick pos=bottom,
        tick align=outside,
        xticklabel style={font=\scriptsize},
        ylabel near ticks,
        xlabel near ticks,
        legend style={at={(0.86,1.4)}, legend columns=2, anchor=north east, nodes={scale=0.9, transform shape},fill=white}, 
        ymajorgrids=true,
        grid style=dashed,
        clip=false
    ]
    \addplot+[
        line width=0.4mm, 
        color=ETHRed, 
        mark=none,
    ] coordinates {
        (0,0)
(1,0)
(2,0)
(3,0)
(4,0)
(5,0)
(6,0)
(7,0)
(8,0)
(9,0)
(10,0)
(11,0)
(12,0)
(13,0)
(14,0)
(15,0)
(16,0)
(17,0)
(18,0)
(19,0)
(20,0)
(21,0)
(22,0)
(23,0)
(24,0)
(25,0)
(26,0)
(27,0)
(28,0)
(29,0)
(30,0)
(31,0)
(32,0)
(33,0)
(34,0)
(35,0)
(36,0)
(37,0)
(38,0)
(39,0)
(40,0)
(41,0)
(42,0)
(43,0)
(44,0)
(45,0)
(46,0)
(47,0)
(48,0)
(49,0)
(50,0)
(51,0)
(52,0)
(53,0)
(54,0)
(55,0)
(56,0)
(57,0)
(58,0)
(59,0)
(60,0)
(61,0)
(62,0)
(63,0)
(64,0)
(65,0)
(66,0)
(67,0)
(68,0)
(69,0)
(70,0)
(71,0)
(72,0)
(73,0)
(74,0)
(75,0)
(76,0)
(77,0)
(78,0)
(79,0)
(80,0)
(81,0)
(82,0)
(83,0)
(84,0)
(85,0)
(86,0)
(87,0)
(88,0)
(89,0)
(90,0)
(91,0)
(92,0)
(93,0)
(94,0)
(95,0)
(96,0)
(97,0)
(98,0)
(99,0)
(100,0)
(101,0)
(102,0)
(103,0)
(104,0)
(105,0)
(106,0)
(107,0)
(108,0)
(109,0)
(110,0)
(111,0)
(112,0)
(113,0)
(114,0)
(115,0)
(116,0)
(117,0)
(118,0)
(119,0)
(120,0)
    };
    \addlegendentry{Vision Gap}
    \addplot+[
        dashed,
        line width=0.4mm, 
        color=ETHBronze, 
        mark=none,
    ] coordinates {
(0, 0.0)
(1, 0.0009765625)
(2, 0.0)
(3, 0.0)
(4, 0.0009765625)
(5, 0.0)
(6, 0.0)
(7, 0.0)
(8, 0.0)
(9, 0.0)
(10, 0.0009765625)
(11, 0.0)
(12, 0.0)
(13, 0.0)
(14, 0.0)
(15, 0.0009765625)
(16, 0.0)
(17, 0.0)
(18, 0.0)
(19, 0.0)
(20, 0.001953125)
(21, 0.0)
(22, 0.0)
(23, 0.0)
(24, 0.0)
(25, 0.0)
(26, 0.0)
(27, 0.0)
(28, 0.0)
(29, 0.0)
(30, 0.00390625)
(31, 0.0)
(32, 0.0)
(33, 0.0)
(34, 0.0)
(35, 0.0)
(36, 0.0)
(37, 0.0)
(38, 0.0)
(39, 0.0)
(40, 0.0068359375)
(41, 0.0)
(42, 0.0)
(43, 0.0)
(44, 0.0)
(45, 0.0)
(46, 0.0)
(47, 0.0)
(48, 0.0)
(49, 0.0)
(50, 10)
(51, 10)
(52, 10)
(53, 10)
(54, 10)
(55, 10)
(56, 10)
(57, 10)
(58, 10)
(59, 10)
(60, 10)
(61, 10)
(62, 10)
(63, 10)
(64, 10)
(65, 10)
(66, 10)
(67, 10)
(68, 10)
(69, 10)
(70, 10)
(71, 10)
(72, 10)
(73, 10)
(74, 10)
(75, 10)
(76, 10)
(77, 10)
(78, 10)
(79, 10)
(80, 10)
(81, 10)
(82, 10)
(83, 10)
(84, 10)
(85, 10)
(86, 10)
(87, 10)
(88, 10)
(89, 10)
(90, 10)
(91, 10)
(92, 10)
(93, 10)
(94, 10)
(95, 10)
(96, 10)
(97, 10)
(98, 10)
(99, 10)
(100,10)
(101, 10)
(102, 10)
(103, 10)
(104, 10)
(105, 10)
(106, 10)
(107, 10)
(108, 10)
(109, 10)
(110, 10)
(111, 10)
(112, 10)
(113, 10)
(114, 10)
(115, 10)
(116, 10)
(117, 10)
(118, 10)
(119, 10)
(120, 10)
    };
    \addlegendentry{Language Gap}
    \draw [decorate, decoration={brace, amplitude=5pt, raise=2pt}, color=ETHBlue, line width=1pt]
    (axis cs:0.5, 10.5) -- (axis cs:48.5, 10.5) 
    node [midway, above=8pt, color=ETHBlue, font=\scriptsize\bfseries] {ID};

    \draw [decorate, decoration={brace, amplitude=5pt, raise=2pt}, color=ETHGreen, line width=1pt]
    (axis cs:50.5, 10.5) -- (axis cs:98.5, 10.5) 
    node [midway, above=8pt, color=ETHGreen, font=\scriptsize\bfseries] {VE};

    \draw [decorate, decoration={brace, amplitude=5pt, raise=2pt}, color=ETHPurple, line width=1pt]
    (axis cs:100.5, 10.5) -- (axis cs:119.5, 10.5) 
    node [midway, above=8pt, color=ETHPurple, font=\scriptsize\bfseries] {FE};

    \draw [dashed, gray, opacity=0.5] (axis cs:49, -2) -- (axis cs:49, 10);
    \draw [dashed, gray, opacity=0.5] (axis cs:99, -2) -- (axis cs:99, 10);

    \end{axis}
    \end{tikzpicture}

    \caption{\textbf{Visual representations remain robust while symbolic decoding collapses.} 
    The \textit{Vision Gap} ($|N_G - N_H|$) remains zero across all regimes, proving the linear probe recovers the true count even during extrapolation. 
    In contrast, the \textit{Language Gap} ($|N_P - N_H|$) explodes immediately when $N \ge 50$. 
    This confirms the bottleneck is not perceptual blindness, but a failure to map valid visual magnitudes to the correct symbolic tokens.}
    \label{fig:syn:recognition}
\end{figure}
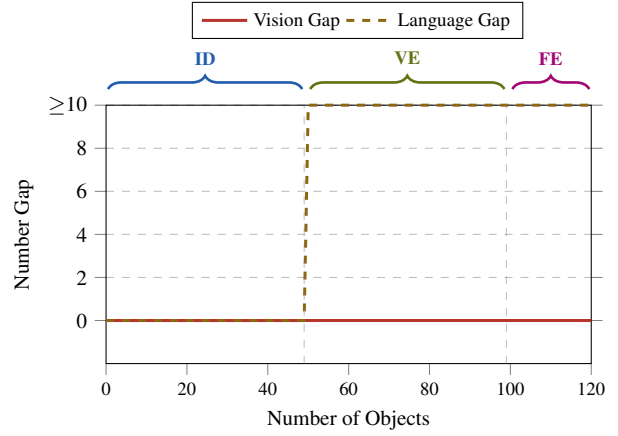
\paragraph{Experiment Results.}
As illustrated in \cref{fig:syn:recognition}, the results reveal a striking dissociation between perception and generation. 
The \textit{Vision Gap} remains effectively zero across the entire spectrum ($0-120$).
The visual encoder maintains a linearly separable representation of quantity well into the FE regime, demonstrating that the visual features for ``120 objects'' are distinct and readable even though the model has never been trained on images with more than 49 objects.
However, the \textit{Language Gap} explodes precisely at the boundary of the training support ($N=50$). 
While the encoder ``sees'' the correct number ($N_H \approx N_G$), the language decoder fails to map the recognized objects to the final generated symbolic token ($N_P \neq N_H$).

\paragraph{Causality Verification.}
To confirm that the probed features are the actual causal drivers of counting rather than spurious correlations, we perform an intervention analysis. 
By masking the latent representations of $k \in [1, 5]$ tokens identified as black stones by our probe (within the ID setting), we observe that the model's prediction drops deterministically to exactly $N_G - k$. 
This confirms a direct causal link between the features isolated by our probe and the model's final counting output. 
Detailed experimental settings and results are provided in \cref{app:sup:results:causal}.

\paragraph{Takeaway.}
We reject \textit{Hypothesis A}.
The visual backbone successfully individuates objects far beyond the training distribution, confirming that the bottleneck is not perceptual.
Since the valid quantity signal ($N_H$) successfully enters the system, the failure must stem from downstream processing rather than blindness.
Having established that the model \textit{sees} the quantity, we proceed to test \textit{Hypothesis B}: does the model (i.e., language decoder) possess a sense of magnitude, or does the signal dissipate during quantity reasoning?

\subsection{Hypothesis B: Latent Magnitude Awareness}
\label{sec:syn:stage2}
To detect if the language module possesses a robust sense of numerosity, we investigate whether the quantity signal persists in its hidden layers and if it can utilize these internal magnitude states for logical discrimination tasks, even when it cannot decode them into specific number tokens.

\paragraph{Experiment Design.}
We evaluate magnitude awareness via a \textit{Comparative Counting} task, reformulating enumeration as a binary classification problem to bypass symbolic articulation bottlenecks.
The model receives two inputs, either two character lists (\textit{Text-to-Text}) or one image and one character list (\textit{Vision-to-Text}), and must generate a boolean token (``True'' or ``False'') indicating if they share the same cardinality (see examples in \cref{fig:prompt_template:syn:text_compare,fig:prompt_template:syn:vision_compare}).
We construct a balanced dataset with equal positive and negative pairs to prevent heuristic guessing.
Crucially, consistent with our counting curriculum (\cref{sec:framework:syn}), the Vision-to-Text training is restricted to visual quantities $N \le 49$, while evaluation extends to the unseen VE regime ($N \in [50, 99]$). 
Success in this regime would prove that the model can map a novel visual magnitude to a known textual magnitude, even if it cannot explicitly name it.

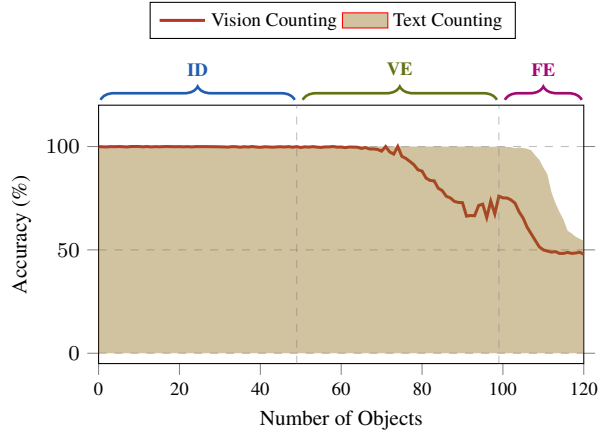
\begin{figure}[!t]
    \centering
    \scalefont{0.8}
    \begin{tikzpicture}
    \begin{axis}[
        width=8cm, height=5cm,
        xlabel={Number of Objects},
        ylabel={Accuracy (\%)},
        xmin=0, xmax=120,
        ymin=-5, ymax=120,
        xtick={0, 20, 40, 60, 80, 100, 120},
        xticklabels={0, 20, 40, 60, 80, 100, 120},
        xtick pos=bottom,
        tick align=outside,
        xticklabel style={font=\scriptsize},
        ylabel near ticks,
        xlabel near ticks,
        legend style={at={(0.86,1.4)}, legend columns=2, anchor=north east, nodes={scale=0.9, transform shape},fill=white},
        ymajorgrids=true,
        grid style=dashed,
        clip=false
    ]

    \addplot+[
        line width=0.4mm, 
        color=ETHRed, 
        mark=none,
    ] coordinates {
(0, 99.951171875)
(1, 99.853515625)
(2, 99.8046875)
(3, 99.951171875)
(4, 99.90234375)
(5, 100.0)
(6, 99.90234375)
(7, 99.8046875)
(8, 100.0)
(9, 100.0)
(10, 100.0)
(11, 99.853515625)
(12, 100.0)
(13, 99.8046875)
(14, 99.951171875)
(15, 99.853515625)
(16, 99.951171875)
(17, 100.0)
(18, 99.951171875)
(19, 99.90234375)
(20, 99.951171875)
(21, 99.853515625)
(22, 99.951171875)
(23, 99.8046875)
(24, 99.951171875)
(25, 99.951171875)
(26, 99.951171875)
(27, 99.951171875)
(28, 99.90234375)
(29, 99.90234375)
(30, 99.853515625)
(31, 99.8046875)
(32, 99.70703125)
(33, 99.951171875)
(34, 99.90234375)
(35, 99.658203125)
(36, 99.90234375)
(37, 99.755859375)
(38, 100.0)
(39, 99.755859375)
(40, 99.609375)
(41, 99.853515625)
(42, 99.8046875)
(43, 99.658203125)
(44, 99.8046875)
(45, 99.90234375)
(46, 99.755859375)
(47, 99.951171875)
(48, 99.658203125)
(49, 99.951171875)
(50, 99.51171875)
(51, 99.853515625)
(52, 99.70703125)
(53, 99.951171875)
(54, 99.70703125)
(55, 99.560546875)
(56, 99.8046875)
(57, 99.8046875)
(58, 100.0)
(59, 99.70703125)
(60, 99.755859375)
(61, 99.462890625)
(62, 99.755859375)
(63, 99.70703125)
(64, 99.658203125)
(65, 99.072265625)
(66, 99.51171875)
(67, 98.828125)
(68, 98.828125)
(69, 98.583984375)
(70, 97.75390625)
(71, 100.0)
(72, 97.36328125)
(73, 96.337890625)
(74, 100.0)
(75, 95.263671875)
(76, 94.23828125)
(77, 92.7734375)
(78, 91.162109375)
(79, 88.720703125)
(80, 88.134765625)
(81, 84.716796875)
(82, 83.59375)
(83, 83.3984375)
(84, 79.833984375)
(85, 78.662109375)
(86, 75.927734375)
(87, 75.146484375)
(88, 73.4375)
(89, 72.94921875)
(90, 72.8515625)
(91, 66.30859375)
(92, 66.650390625)
(93, 66.50390625)
(94, 71.484375)
(95, 72.119140625)
(96, 65.673828125)
(97, 73.583984375)
(98, 67.431640625)
(99, 75.9765625)
(100, 75.146484375)
(101, 75.09765625)
(102, 74.169921875)
(103, 72.607421875)
(104, 68.408203125)
(105, 65.380859375)
(106, 60.9375)
(107, 57.763671875)
(108, 54.58984375)
(109, 51.46484375)
(110, 50.0)
(111, 49.4140625)
(112, 48.92578125)
(113, 49.169921875)
(114, 48.291015625)
(115, 48.291015625)
(116, 48.779296875)
(117, 48.291015625)
(118, 48.486328125)
(119, 48.876953125)
(120, 47.900390625)
    };
    \addlegendentry{Vision Counting}

    \addplot+[
        draw=none,          %
        fill=ETHBronze,     %
        fill opacity=0.4,   %
        mark=none,
        area legend
    ] coordinates {
(0, 99.90234375)
(1, 99.755859375)
(2, 99.853515625)
(3, 99.8046875)
(4, 99.8046875)
(5, 99.8046875)
(6, 99.853515625)
(7, 99.90234375)
(8, 99.853515625)
(9, 100.0)
(10, 100.0)
(11, 100.0)
(12, 99.951171875)
(13, 99.951171875)
(14, 99.951171875)
(15, 99.90234375)
(16, 99.951171875)
(17, 99.90234375)
(18, 99.951171875)
(19, 100.0)
(20, 100.0)
(21, 99.951171875)
(22, 100.0)
(23, 100.0)
(24, 99.90234375)
(25, 100.0)
(26, 100.0)
(27, 100.0)
(28, 100.0)
(29, 100.0)
(30, 100.0)
(31, 99.951171875)
(32, 100.0)
(33, 100.0)
(34, 100.0)
(35, 100.0)
(36, 100.0)
(37, 100.0)
(38, 100.0)
(39, 100.0)
(40, 99.951171875)
(41, 100.0)
(42, 100.0)
(43, 100.0)
(44, 100.0)
(45, 100.0)
(46, 100.0)
(47, 100.0)
(48, 100.0)
(49, 100.0)
(50, 100.0)
(51, 99.951171875)
(52, 100.0)
(53, 100.0)
(54, 100.0)
(55, 100.0)
(56, 100.0)
(57, 100.0)
(58, 100.0)
(59, 100.0)
(60, 100.0)
(61, 100.0)
(62, 100.0)
(63, 100.0)
(64, 99.951171875)
(65, 100.0)
(66, 100.0)
(67, 100.0)
(68, 100.0)
(69, 100.0)
(70, 100.0)
(71, 100.0)
(72, 100.0)
(73, 99.951171875)
(74, 100.0)
(75, 100.0)
(76, 100.0)
(77, 100.0)
(78, 100.0)
(79, 99.90234375)
(80, 100.0)
(81, 100.0)
(82, 100.0)
(83, 100.0)
(84, 100.0)
(85, 100.0)
(86, 99.951171875)
(87, 100.0)
(88, 99.951171875)
(89, 99.951171875)
(90, 100.0)
(91, 100.0)
(92, 99.951171875)
(93, 99.951171875)
(94, 100.0)
(95, 99.951171875)
(96, 99.90234375)
(97, 99.853515625)
(98, 99.951171875)
(99, 99.951171875)
(100, 99.90234375)
(101, 99.8046875)
(102, 99.70703125)
(103, 99.169921875)
(104, 99.560546875)
(105, 99.267578125)
(106, 98.92578125)
(107, 98.193359375)
(108, 95.703125)
(109, 93.65234375)
(110, 89.84375)
(111, 86.62109375)
(112, 77.9296875)
(113, 72.607421875)
(114, 68.310546875)
(115, 64.599609375)
(116, 59.27734375)
(117, 58.0078125)
(118, 56.34765625)
(119, 55.37109375)
(120, 54.296875)
    }\closedcycle;
    \addlegendentry{Text Counting}

    \draw [decorate, decoration={brace, amplitude=5pt, raise=2pt}, color=ETHBlue, line width=1pt]
    (axis cs:0.5, 120) -- (axis cs:48.5, 120) 
    node [midway, above=8pt, color=ETHBlue, font=\scriptsize\bfseries] {ID};

    \draw [decorate, decoration={brace, amplitude=5pt, raise=2pt}, color=ETHGreen, line width=1pt]
    (axis cs:50.5, 120) -- (axis cs:98.5, 120) 
    node [midway, above=8pt, color=ETHGreen, font=\scriptsize\bfseries] {VE};

    \draw [decorate, decoration={brace, amplitude=5pt, raise=2pt}, color=ETHPurple, line width=1pt]
    (axis cs:100.5, 120) -- (axis cs:119.5, 120) 
    node [midway, above=8pt, color=ETHPurple, font=\scriptsize\bfseries] {FE};

    \draw [dashed, gray, opacity=0.5] (axis cs:49, -5) -- (axis cs:49, 120);
    \draw [dashed, gray, opacity=0.5] (axis cs:99, -5) -- (axis cs:99, 120);

    \end{axis}
    \end{tikzpicture}
    \caption{\textbf{Models can compare quantities they cannot name.} 
    In the binary magnitude comparison task, the model maintains robust accuracy ($>70\%$) for visual inputs throughout the VE regime ($50-99$), despite having $0\%$ accuracy in explicit enumeration. 
    This proves the model possesses robust magnitude awareness; the failure is isolated strictly to the generative decoding stage.}
    \label{fig:syn:quantity:train}
\end{figure}

\paragraph{Experiment Results.}
As shown in \cref{fig:syn:quantity:train}, the model exhibits remarkable latent robustness.
Despite achieving $0\%$ accuracy in explicitly \textit{articulating} quantities in the VE regime ($50-99$), it successfully performs cross-modal comparison, e.g., correctly matching an image of 75 stones to a text sequence of 75 characters, with high accuracy.
This performance gradually starts to degrade in the middle of VE regime and degrade to random guessing in FE regime, where neither visual density nor textual quantity was observed.
To further validate this, we provide auxiliary linear probing results in \cref{app:sup:results:syn_probe}, which confirm that linearly separable magnitude representations are preserved with $100\%$ fidelity throughout the language decoder layers.

\paragraph{Takeaway.}
We reject \textit{Hypothesis B}. 
The model possesses robust \textit{magnitude awareness}, successfully utilizing abstract cardinality for quantity reasoning well beyond its training distribution. 
This proves that the quantity signal does not dissipate during cognitive processing.
With perception (Stage 1) and numerosity (Stage 2) cleared, the bottleneck is isolated to the final stage. 
We thus proceed to investigate \textit{Hypothesis C}: a failure in symbolic mapping.

\subsection{Hypothesis C: Symbolic Mapping}
\label{sec:syn:stage3}

Having rejected perceptual failure (\cref{sec:syn:stage1}) and quantity reasoning loss (\cref{sec:syn:stage2}), we turn to \textit{Hypothesis C (fractured magnitude hypothesis)}: a structural failure in projecting the valid magnitude onto the correct symbolic token. 
We investigate this through two complementary analyses: first examining the \textit{output topology} to see \textit{how} it fails, and second examining the \textit{internal circuits} to understand \textit{why}.

\subsubsection{Analysis I: Error Topology}
\paragraph{Experiment Design.}
We verify if the mapping failure is stochastic (random noise) or structural (systematic misalignment). 
We collect the model's predictions ($N_P$) across the entire extrapolation set and analyze the error distribution.
If the model suffers from uncertainty, errors should be around the ground truth. 
If the generative mapping path is broken, we expect the model to collapse to specific statistical priors or ``attractors'' from its training distribution.

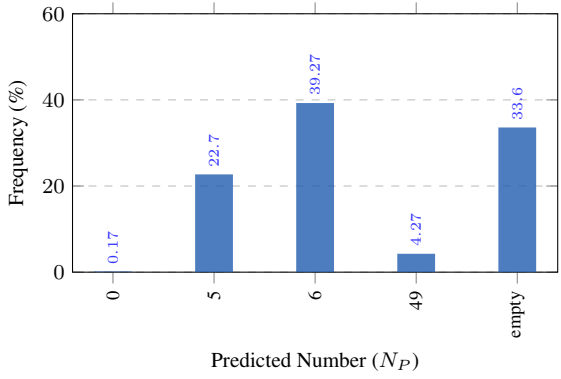
\begin{figure}[!t]
    \centering
    \scalefont{0.8}
    \begin{tikzpicture}
    \begin{axis}[
        ybar,
        width=8cm, height=5cm, %
        xlabel={Predicted Number ($N_P$)},
        ylabel={Frequency (\%)},
        ymin=0, ymax=60, %
        symbolic x coords={0, 5, 6, 49, empty},
        xtick=data,
        nodes near coords,
        nodes near coords align={vertical},
        every node near coord/.append style={font=\tiny, rotate=90, anchor=west}, 
        xticklabel style={font=\scriptsize, rotate=90},
        ylabel near ticks,
        xlabel near ticks,
        bar width=0.5cm, %
        enlarge x limits=0.1,
        ymajorgrids=true,
        grid style=dashed,
    ]

    \addplot+[
        fill=ETHBlue, 
        draw=none,
        opacity=0.8
    ] coordinates {
        (0, 0.166015625)
        (5, 22.701171875)
        (6, 39.267578125)
        (49, 4.26953125)
        (empty, 33.595703125)
    };

    \end{axis}
    \end{tikzpicture}
    \caption{\textbf{Decoding failure manifests as mode collapse.} 
    Instead of random noise, prediction errors cluster tightly around specific ``attractor'' tokens: the visual training boundary ($49$), 
    small-number hallucinations ($5$, $6$), and empty outputs.
    This confirms that in the absence of a valid cross-modal path, the model abandons the counting algorithm and reverts to retrieving memorized priors.}
    \label{fig:syn:label:case}
\end{figure}

\paragraph{Experiment Results.}
As shown in \cref{fig:syn:label:case}, the error distribution is highly non-random. 
Rather than exhibiting Gaussian noise, the model collapses to specific "attractor" tokens: the boundary of the visual training data ($49$), small-number hallucinations (e.g., $5$, $6$), or generating an empty output (EOS).
This confirms that in the absence of a learned mapping path, the model abandons the counting algorithm entirely, retrieving memorized statistical priors instead.

\paragraph{Takeaway.}
The non-Gaussian nature of the errors suggests that the model is not merely ``unsure''; it is fundamentally ungrounded. 
When the input magnitude ($N_H$) falls into the extrapolation regime, the projection layer $f(N_H) \rightarrow N_P$ fails to trigger, causing the model to default to known high-probability tokens.

\subsubsection{Analysis II: Circuit Divergence}

\paragraph{Experiment Design.}
To determine if this grounding failure stems from physical circuit isolation, we trace the activations of the language module's attention heads.
We employ a greedy pruning strategy to identify ``critical heads'': the minimal set of attention heads required to maintain prediction accuracy.
We sequentially drop heads (masking their output) from top to bottom layers and measure the performance drop.
We then calculate the intersection of these critical head sets between the \textit{visual counting} task and the \textit{textual Ccounting} task to quantify their functional overlap.

\begin{figure}[!t]
    \centering
    \scalefont{0.8}
    \pgfplotsset{
        colormap={activations}{color(0cm)=(white); color(1cm)=(ETHRed)}
    }
    \pgfplotsset{
        heatmap style/.style={
            width=3.5cm, height=3.5cm,
            view={0}{90},     %
            xlabel={Head Index},
            ylabel={Layer Index},
            xmin=0.5, xmax=2.5,
            ymin=0.5, ymax=2.5,
            xtick={1, 2},
            ytick={1, 2},
            xticklabel style={font=\scriptsize},
            yticklabel style={font=\scriptsize},
            ylabel near ticks,
            xlabel near ticks,
        }
    }
    \begin{subfigure}[t]{0.48\linewidth}
        \centering
        \begin{tikzpicture}
        \begin{axis}[
            heatmap style,
        ]
        \addplot3[
            matrix plot*,
            mesh/rows=2,
            mesh/cols=2,
            draw=gray!20
        ] coordinates {
            (1,1,0.841640625) (2,1,0.935)
            
            (1,2,0.6128125) (2,2,0.894765625)
        };
        \end{axis}
        \end{tikzpicture}
        \caption{Visual Counting}
    \end{subfigure}%
    \begin{subfigure}[t]{0.48\linewidth}
        \centering
        \begin{tikzpicture}
        \begin{axis}[
            heatmap style,
            colorbar, %
            colorbar style={
                title style={yshift=2pt},
                yticklabel style={font=\tiny},
                width=0.2cm,
            }
        ]
        \addplot3[
            matrix plot*,
            mesh/rows=2,
            mesh/cols=2,
            draw=gray!20
        ] coordinates {
            (1,1,0.8371875) (2,1,0.91953125)
            (1,2, 0.8309375) (2,2,0.75359375)
        };
        \end{axis}
        \end{tikzpicture}
        \caption{Textual Counting}
    \end{subfigure}
    \begin{subfigure}[t]{.96\linewidth}
        \centering
        \begin{tikzpicture}
        \begin{axis}[
            ybar,
            width=8cm, height=5cm,
            ylabel={Frequency (\%)},
            ymin=0, ymax=100,
            symbolic x coords={Same Module Set, Different Module Set},
            xtick={Same Module Set, Different Module Set},
            nodes near coords align={vertical},
            every node near coord/.append style={font=\tiny,}, 
            xticklabel style={font=\scriptsize},
            ylabel near ticks,
            tick style={draw=none},
            bar width=1.5cm,
            enlarge x limits=.7,
            ymajorgrids=true,
            grid style=dashed,
        ]

        \addplot+[
            fill=ETHBlue,
            draw=none,
            opacity=0.8,
            bar shift=0pt,
        ] coordinates {
            (Same Module Set, 26.42)
        };
        
        \addplot+[
            fill=ETHGreen,
            draw=none,
            opacity=0.8,
            bar shift=0pt,
        ] coordinates {
            (Different Module Set, 73.58)
        };
        \end{axis}
        \end{tikzpicture}
        \caption{Activation Pattern Overlap ($N=6400$ Examples)}
    \end{subfigure}
    \caption{\textbf{Visual and textual counting rely on disjoint neural circuits.} 
    \textbf{Top:} Attention heatmaps reveal distinct activation patterns for visual (left) vs. textual (right) counting. 
    \textbf{Bottom:} Quantitative analysis ($N=6400, 12.5\%$ test data) confirms that the set of critical attention heads is non-overlapping in $73.58\%$ of cases. 
    This provides mechanistic evidence for the \textit{Fractured Magnitude Hypothesis}: the model lacks a unified number space, processing visual and textual quantities on disjoint statistical manifolds.}
    \label{fig:syn:label:module}
\end{figure}
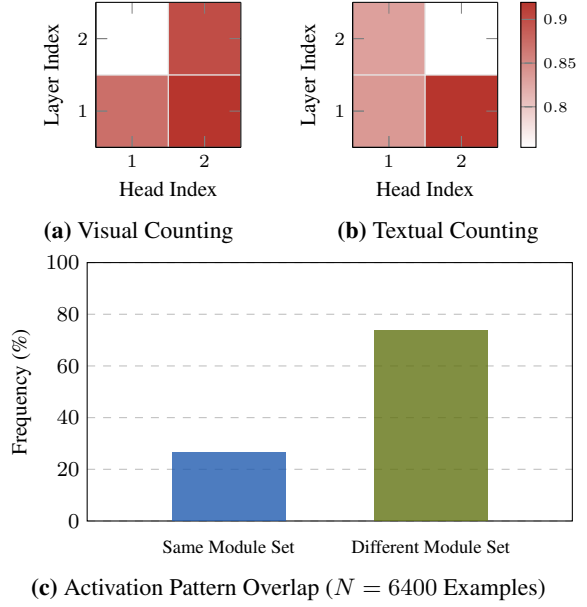

\paragraph{Experiment Results.}
The heatmap analysis (\cref{fig:syn:label:module}) uncovers a startling separation: visual counting and textual counting activate almost entirely disjoint sets of attention heads. 
In $73.58\%$ of cases, the neural circuits responsible for processing quantity in vision do not overlap with those used for text.
This indicates that the model is effectively running two different sub-programs depending on the input modality, with no shared ``counting circuit'' to bridge them.

\paragraph{Takeaway.}
These results confirm \textit{Hypothesis C (fractured magnitude hypothesis)}.
The VLM has not learned a \textit{universal number space}. 
Instead, it has acquired two separate, modality-specific statistical manifolds: a ``visual magnitude'' and a ``textual magnitude''.
Because these manifolds reside in disjoint subspaces (evidenced by the disjoint attention heads), the model cannot generalize the mapping function to unseen quantities. 
While perception generalizes perfectly and magnitude awareness generalizes robustly, the symbolic mapping link is fractured. 
The ``concept of 75'' in vision has no neural bridge to the ``concept of 75'' in text.

\section{Real-World Interpolation Validation}
\label{sec:real}
To verify whether the bottleneck observed in our synthetic laboratory generalizes to large-scale, pretrained architectures, we evaluate Qwen3-VL. 
Analyzing this state-of-the-art foundation model allows us to test if trillion-token pretraining and RLHF alignment are sufficient to overcome the \textit{fractured magnitude} phenomenon.

\subsection{Persistence of the Extrapolation Gap}
\label{sec:real:acc}

We first validate if the ``baseline paradox'' persists in a foundation model. 
If Qwen3-VL exhibits the same dissociation between textual and visual counting, it suggests the failure is structural rather than an artifact of scale.

\paragraph{Experiment Design.}
We adopt the evaluation protocol from \cref{sec:syn:acc}, assessing both \textit{visual counting} (black stones) and \textit{textual counting} (characters). 
The only modification is the dataset: we use $6\times6$ Go boards to align with the model's effective resolution capacity.

\begin{figure}[!t]
    \centering
    \scalefont{0.8}
    \begin{tikzpicture}
    \begin{axis}[
        width=8cm, height=5.cm,
        xlabel={Number of Objects},
        ylabel={Accuracy (\%)},
        xmin=0, xmax=20,
        ymin=-5, ymax=120,
        xtick={0, 4, 8, 12, 16, 20},
        xticklabels={0, 4, 8, 12, 16, 20},
        xtick pos=bottom,
        tick align=outside,
        xticklabel style={font=\scriptsize},
        ylabel near ticks,
        xlabel near ticks,
        legend style={at={(0.96,0.95)}, legend columns=1, anchor=north east, nodes={scale=0.9, transform shape},fill=white}, 
        ymajorgrids=true,
        grid style=dashed,
        clip=false
    ]

    \addplot+[
        solid,
        line width=0.4mm,
        color=ETHRed, 
        mark=none,
    ] coordinates {
(0, 100.0)
(1, 100.0)
(2, 94.0)
(3, 78.0)
(4, 62.0)
(5, 56.00000000000001)
(6, 43.0)
(7, 33.0)
(8, 28.000000000000004)
(9, 28.000000000000004)
(10, 21.0)
(11, 17.0)
(12, 18.0)
(13, 10.0)
(14, 6.0)
(15, 6.0)
(16, 3.0)
(17, 3.0)
(18, 16.0)
(19, 6.0)
(20, 4.0)
    };
    \addlegendentry{Vision Counting}

    \addplot+[
        draw=none,          %
        fill=ETHBronze,     %
        fill opacity=0.4,   %
        mark=none,
        area legend
    ] coordinates {
(0, 100.0)
(1, 100.0)
(2, 100.0)
(3, 100.0)
(4, 100.0)
(5, 98.0)
(6, 91.0)
(7, 80.0)
(8, 64.0)
(9, 62.0)
(10, 56.99999999999999)
(11, 50.0)
(12, 53.0)
(13, 44.0)
(14, 49.0)
(15, 49.0)
(16, 49.0)
(17, 56.00000000000001)
(18, 63.0)
(19, 56.99999999999999)
(20, 56.99999999999999)
    }\closedcycle;
    \addlegendentry{Text Counting}
    \draw [dashed, gray, opacity=0.5] (axis cs:4, -5) -- (axis cs:4, 120);
    \draw [dashed, gray, opacity=0.5] (axis cs:8, -5) -- (axis cs:8, 120);

    \end{axis}
    \end{tikzpicture}
    \caption{\textbf{The extrapolation gap persists in foundation models.} 
    While Text Counting remains robust ($>50\%$), Vision Counting collapses immediately after the subitizing range ($N>5$), approaching zero by $N=15$. 
    This confirms that even trillion-token scale models fail to ground their algorithmic counting rules visually, treating the modalities as disjoint capabilities.}
    \label{fig:real:acc}
\end{figure}
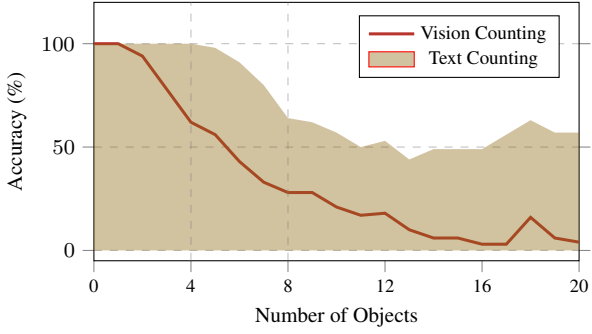
\paragraph{Experiment Results.}
As shown in \cref{fig:real:acc}, the results strikingly mirror our synthetic findings. 
Textual counting remains robust ($100\%$ accuracy on small numbers), whereas visual counting exhibits a precipitous decline, flatlining near random guessing for $N > 8$. 
This confirms that the counting bottleneck is structural; linguistic fluency does not automatically ground visual quantities.

\paragraph{Takeaway.}
We confirm that the \textit{baseline paradox} is a structural characteristic of VLM architectures, not merely an artifact of scale. 
Despite trillion-token pretraining, the model exhibits the same dissociation: algorithmic competence in text versus stochastic failure in vision. 
Having established this baseline, we proceed to verify if the mechanism of failure follows the same diagnostic path (Hypotheses A $\rightarrow$ C) as our synthetic laboratory.

\subsection{Hypothesis A: Object Individuation}
\label{sec:real:stage1}

A common critique of VLMs is that they suffer from ``perceptual crowding'': an inability to distinguish individual features in complex real-world encoders (e.g., SigLIP or CLIP) compared to cleaner synthetic encoders. 
We investigate whether Qwen3-VL's failure to count stems from being effectively ``blind'' to the objects, or if the \textit{hidden number} ($N_H$) persists in the latent space despite the incorrect output.

\paragraph{Experiment Design.}
We replicate the \textit{hidden number probing} methodology from \cref{sec:syn:stage1}. 
Specifically, we train a linear regression probe on the feature map outputs to predict object density per token and aggregate them to estimate the hidden quantity $N_H$.\footnote{Note that the hidden state dimension in Qwen3-VL is quite large. We preprocess it by embedding it into 32-dimensional space.} 
We then compare two error metrics: the \textit{Vision Gap} and the \textit{Language Gap}.

\begin{figure}[!t]
    \centering
    \scalefont{0.8}
    \begin{tikzpicture}
    \begin{axis}[
        width=8cm, height=5.cm,
        xlabel={Number of Objects},
        ylabel={Number Gap},
        xmin=0, xmax=20,
        ymin=-2, ymax=10,
        xtick={0, 4, 8, 12, 16, 20},
        xticklabels={0, 4, 8, 12, 16, 20},
        ytick={0, 2, 4, 6, 8},
        yticklabels={0, 2, 4, 6, 8},
        xtick pos=bottom,
        tick align=outside,
        xticklabel style={font=\scriptsize},
        ylabel near ticks,
        xlabel near ticks,
        legend style={at={(0.42,0.95)}, anchor=north east, nodes={scale=0.9, transform shape}, fill=white,opacity=0.8}, 
        ymajorgrids=true,
        grid style=dashed,
        clip=false
    ]
    \addplot+[
        line width=0.4mm, 
        color=ETHRed, 
        mark=none,
    ] coordinates {
(0, 0.0)
(1, 0.0)
(2, 0.011111111111111112)
(3, 0.0)
(4, 0.0)
(5, 0.0)
(6, 0.0)
(7, 0.0)
(8, 0.0)
(9, 0.0)
(10, 0.0)
(11, 0.0)
(12, 0.0)
(13, 0.0)
(14, 0.0)
(15, 0.0)
(16, 0.0)
(17, 0.0)
(18, 0.022222222222222223)
(19, 0.022222222222222223)
(20, 0.07777777777777778)
    };
    \addlegendentry{Vision Gap}
    \addplot+[
        dashed,
        line width=0.4mm, 
        color=ETHBronze, 
        mark=none,
    ] coordinates {
(0, 0.0)
(1, 0.0)
(2, 0.044444444444444446)
(3, 0.25555555555555554)
(4, 0.5666666666666667)
(5, 0.7222222222222222)
(6, 1.1444444444444444)
(7, 1.5333333333333334)
(8, 1.6555555555555554)
(9, 2.0555555555555554)
(10, 2.522222222222222)
(11, 2.988888888888889)
(12, 2.9)
(13, 4.455555555555556)
(14, 4.411111111111111)
(15, 5.6)
(16, 5.355555555555555)
(17, 4.588888888888889)
(18, 3.911111111111111)
(19, 3.7111111111111112)
(20, 4.2)
    };
    \addlegendentry{Language Gap}
    \end{axis}
    \end{tikzpicture}
    \caption{\textbf{Visual encoders retain rich signal.} 
    The Vision Gap ($|N_G - N_H|$) remains negligible ($\!<\!0.1$), indicating the probe recovers the true count even when the model fails. 
    In contrast, the Language Gap ($|N_P - N_H|$) diverges rapidly. 
    This confirms that while real-world encoders introduce minor noise, the primary driver of failure remains the symbolic decoding bottleneck.}
    \label{fig:real:recognition}
\end{figure}
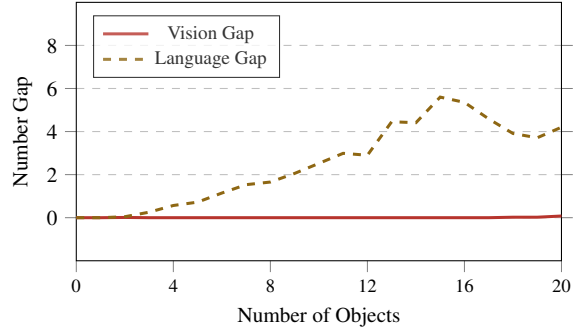

\paragraph{Experiment Results.}
The results in \cref{fig:real:recognition} reveal a crucial distinction between noise and failure. 
Unlike the synthetic setting where recognition was perfectly lossless, the real-world encoder exhibits a non-zero Vision Gap, reflecting the inherent noise of processing high-dimensional natural image features. 
However, this error is negligible ($<0.1$ on average) compared to the Language Gap, which diverges rapidly as soon as the quantity exceeds the immediate subitizing range ($N > 4$). 
The probe successfully tracks the hidden number ($N_H \approx N_G$), proving that the deviation in the final prediction is driven almost entirely by the language module failure ($|N_H - N_P|$).
It is worth noting that while our controlled setting isolates numerosity, real-world counting tasks often involve complex object recognition challenges (e.g., occlusion, variance) which may introduce additional perceptual errors not captured here (\cref{sec:limitations}).

\paragraph{Takeaway.}
We reject \textit{Hypothesis A}. 
Consistent with our synthetic findings, the visual encoder is not the bottleneck. 
While real-world features introduce minor perceptual noise, the \textit{hidden number} remains distinct and linearly separable ($N_H \approx N_G$). 
The catastrophic drop in performance is not due to blindness; the model successfully encodes the visual quantity but fails to utilize it.

\subsection{Hypothesis B: Magnitude Awareness}
\label{sec:real:stage2}

We next investigate the process in language decoder. Does the abstract ``sense'' of quantity dissipate as information propagates through the deep transformer layers? 
We probe the network's depth to determine if the counting failure stems from ``awareness drift'' or catastrophic forgetting within the language module backbone.

\paragraph{Experiment Design.}
For visual counting, we train linear probes on the output hidden states of every layer in the architecture: from the final layer of the vision encoder through the first 63 layers of the language decoder.\footnote{We exclude the last layer as it is directly for logit projection.}
This allows us to trace the fidelity of the magnitude signal ($N_H$) as it flows through the network's computation graph.

\begin{figure}[!t]
    \centering
    \scalefont{0.8}
    \begin{tikzpicture}
    \begin{axis}[
        width=8cm, height=5.cm,
        xlabel={Layer Index},
        ylabel={Probe Accuracy (\%)},
        xmin=-2, xmax=66,
        ymin=0, ymax=110, %
        xtick={0, 10, 20, 30, 40, 50, 60},
        xticklabels={0, 10, 20, 30, 40, 50, 60},
        ylabel near ticks,
        xlabel near ticks,
        ymajorgrids=true,
        grid style=dashed,
        legend style={at={(0.53,0.5)}, anchor=north east, nodes={scale=0.9, transform shape}, fill=white},
    ]

    \fill[ETHBlue!10] (axis cs:-2,0) rectangle (axis cs:0.5,110);
    \fill[ETHGreen!10] (axis cs:0.5,0) rectangle (axis cs:66,110);

    \addplot+[
        only marks,
        mark=*,
        mark size=3.5pt,
        color=ETHBlue,
        mark options={fill=ETHBlue}
    ] coordinates {
        (0, 99.37)
    };
    \addlegendentry{Visual Module}

    \addplot+[
        thick,
        color=ETHGreen,
        mark=none, 
        smooth
    ] coordinates {
        (0, 99.36507936507937)
(1, 100.0)
(2, 99.7883597883598)
(3, 99.7883597883598)
(4, 99.7883597883598)
(5, 99.7883597883598)
(6, 99.7883597883598)
(7, 99.7883597883598)
(8, 99.7883597883598)
(9, 99.7883597883598)
(10, 99.7883597883598)
(11, 99.7883597883598)
(12, 99.7883597883598)
(13, 99.84126984126985)
(14, 100.0)
(15, 100.0)
(16, 99.94708994708995)
(17, 99.94708994708995)
(18, 99.94708994708995)
(19, 100.0)
(20, 100.0)
(21, 100.0)
(22, 100.0)
(23, 100.0)
(24, 100.0)
(25, 100.0)
(26, 100.0)
(27, 100.0)
(28, 100.0)
(29, 100.0)
(30, 100.0)
(31, 99.94708994708995)
(32, 100.0)
(33, 99.94708994708995)
(34, 99.89417989417989)
(35, 99.94708994708995)
(36, 99.94708994708995)
(37, 99.84126984126985)
(38, 98.99470899470899)
(39, 99.1005291005291)
(40, 97.72486772486772)
(41, 97.03703703703704)
(42, 97.19576719576719)
(43, 96.71957671957672)
(44, 95.34391534391534)
(45, 95.92592592592592)
(46, 90.42328042328043)
(47, 88.57142857142857)
(48, 85.71428571428571)
(49, 85.97883597883597)
(50, 86.03174603174604)
(51, 85.76719576719577)
(52, 83.96825396825398)
(53, 80.47619047619048)
(54, 76.19047619047619)
(55, 73.22751322751323)
(56, 71.05820105820105)
(57, 68.994708994709)
(58, 68.2010582010582)
(59, 68.78306878306879)
(60, 67.77777777777779)
(61, 65.23809523809524)
(62, 65.13227513227513)
    };
    \addlegendentry{LLM Layers (1-62)}
    \addplot+[
        only marks,
        mark=*,
        mark size=3.5pt,
        color=ETHBlue,
        mark options={fill=ETHGreen}
    ] coordinates {
        (63, 65.60846560846561)
    };
    \addlegendentry{Language Module}

    \draw [dashed, thick, gray] (axis cs:0.5, 0) -- (axis cs:0.5, 110);

    \end{axis}
    \end{tikzpicture}
    \caption{\textbf{Magnitude information persists deep into the model.} 
    Probing reveals that visual module ($99.4\%$) and the first 40 language module layers ($>99\%$) maintain precise quantity representations. 
    The signal decay is restricted to the final layers, confirming that the failure is not due to ``reasoning drift'' or forgetting, but occurs strictly during the final projection to symbolic logits.}
    \label{fig:real:quantity:probe}
\end{figure}
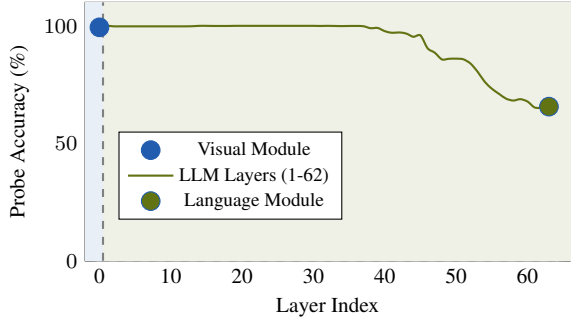 
\paragraph{Experiment Results.}
\cref{fig:real:quantity:probe} presents the probe accuracy across the network depth. 
The results are unequivocal: magnitude information is robustly preserved for the vast majority of the computation.
The visual module passes a precise signal ($99.37\%$) to the language model. 
Remarkably, the language module maintain this fidelity with near-perfect accuracy ($>99\%$) for the first 40 layers. 
We only observe a mild degradation in the final 20 layers, corresponding to the transition from abstract magnitude awareness to the specific (and fractured) output token space.

\paragraph{Takeaway.}
We reject \textit{Hypothesis B}. 
The model effectively ``knows'' the cardinality of the scene throughout its decoder. 
The fact that accuracy remains at ceiling levels for over two-thirds of the network depth rules out magnitude awareness capacity or forgetting as the primary bottleneck. 
While the mild degradation in the final layers suggests some noise is introduced as the model prepares to generate output, this marginal error is insufficient to explain the catastrophic collapse in visual counting performance. 
This indicates that the failure is not one of intellect, but of articulation: the final \textit{symbolic mapping} could be broken.

\subsection{Hypothesis C: Symbolic Mapping}
\label{sec:real:stage3}
While \cref{sec:real:acc} proves the model can count algorithmically in text, accuracy metrics alone are insufficient to reveal the mechanism of the visual failure.
Does the model attempt to run this same counting algorithm on visual inputs (suffering merely from execution noise), or does it abandon the procedure entirely in favor of ungrounded estimation?

\paragraph{Experiment Design.}
We apply the \textit{error topology analysis} from \cref{sec:syn:stage3}. 
We diagnose the failure mechanism by contrasting the statistical signatures of \textit{variance} (spread) and \textit{bias} (offset) in the prediction heatmaps ($N_P$ vs. $N_G$).
A tight clustering with consistent bias (e.g., $N \pm 1$) indicates \textit{algorithmic execution}, implying the rule is active but misaligned.
In contrast, a diffuse, cloud-like spread indicates \textit{heuristic estimation}, implying the model has reverted to ungrounded guessing.

\input{figures/results/real_label_case}

\paragraph{Experiment Results.}
\cref{fig:real:label:case} illustrates a stark dichotomy in error topology. 
\textit{Visual counting} errors form a diffuse, high-variance cloud, where the model frequently over- or under-estimates by wide margins. 
The mean prediction (Red Line) drifts significantly from the ground truth diagonal, a pattern characteristic of \textit{approximate estimation} rather than precise enumeration.
In sharp contrast, \textit{textual counting} exhibits errors that are tightly clustered along the diagonal. 
The model often predicts exactly $N \pm 1$, maintaining the correct slope but suffering from a constant offset. 
This low-variance, systematic bias indicates a working \textit{algorithm estimation} with a calibration error (likely due to tokenization artifacts, e.g., counting the start-of-sentence token), rather than a fundamental loss of magnitude awareness.

\paragraph{Takeaway.}
These results confirm \textit{Hypothesis C} in a foundation model setting.
The analysis of Qwen3-VL validates the \textit{fractured magnitude hypothesis} in a real-world setting.
The model essentially operates with two distinct cognitive processes: a \textit{robust textual counter} that executes a precise (albeit biased) algorithm, and a \textit{stochastic visual estimator} that relies on ungrounded heuristics.
Crucially, the model possesses both a precise visual magnitude (proven by probing) and a systematic counting algorithm (proven by accuracy), yet it fails to bridge them. 
The \textit{universal number space} required to map the visual representation to the textual token is missing, forcing the model to guess stochastically despite having the correct answer encoded in its latent space.

\section{Conclusion}
\label{sec:conclusion}
We diagnose the bottleneck in visual counting, moving from the controlled synthetic laboratory to the state-of-the-art foundation model.
Our three-stage decomposition rules out failures in \textit{visual individuation} and \textit{magnitude awareness}: models maintain robust latent quantities and support cross-modal comparison even when generation fails.
Instead, the collapse is strictly confined to \textit{symbolic mapping}.
This supports a \textit{fractured magnitude hypothesis}: disjoint visual and textual manifolds prevent the grounding of unseen quantities.
Consequently, data scaling is insufficient; systematic generalization requires architectural priors that enforce unified, isomorphic number spaces across modalities.

\section*{Acknowledgment}
We thank the reviewers for their constructive feedback. 
We also thank Jiaoda Li and Mubashara Akhtar for their valuable input during the early stages of this work.
Yifan Hou is supported by the Swiss Data Science Center PhD Grant (P22-05). This research was further supported by the ETH AI Center
through an ETH AI Center doctoral fellowship to Junling Wang.

\section*{Impact Statement}
This paper presents foundational research aimed at understanding the systematic generalization capabilities and failure modes of Multimodal Large Language Models (e.g., VLMs). By diagnosing the ``Fractured Magnitude'' bottleneck, our work highlights a critical discrepancy between a model's internal representation (which is often accurate) and its generated output (which can be hallucinated).

\paragraph{Societal Consequences.} 
This work focuses on \textit{mechanistic interpretability} and does not directly introduce new applications with immediate dual-use risks. 
However, improving the counting and reasoning reliability of VLMs has positive downstream implications for scientific analysis, automated inventory, and data processing tasks where precision is critical. 
Conversely, as with all advancements in reasoning capabilities, reliable VLMs could theoretically be employed in autonomous surveillance systems; however, our work identifies current limitations rather than enabling immediate deployment capabilities.

\paragraph{Environmental Impact.} 
Our study utilizes a hybrid approach, leveraging lightweight synthetic models (trained from scratch) to derive hypotheses before validating them on pre-trained foundation models. This methodology significantly reduces the carbon footprint compared to studies that require full-scale retraining of large foundation models.

We do not foresee any specific negative societal consequences or ethical concerns arising directly from the publication of this analysis.

\bibliography{example_paper}
\bibliographystyle{icml2026}

\newpage
\appendix
\onecolumn

\section{Related Work}
\label{sec:related}

\subsection{Visual Counting and Numerosity}
The task of estimating the number of objects in a visual scene has evolved from task-specific engineering to general-purpose reasoning. 
Traditional approaches treated counting as a density estimation or regression problem, utilizing dedicated architectures like U-Nets or crowd-counting density map regressors to handle high-density scenes~\cite{learntocount_lempitsky_nips2010,singleimg_zhang_cvpr2016,unet_ronneberger_miccai2015,falk2019unet}. 
While highly accurate, these models are restricted to specific categories (e.g., people, cars) and lack semantic flexibility.
The advent of Vision-Language Models (VLMs) introduced the promise of open-vocabulary zero-shot counting~\cite{clipcount_jiang_mm2023,clipcount_paiss_iccv2023,zscount_xu_cvpr2023}. 
However, recent evaluations consistently demonstrate that while VLMs excel at object recognition, they struggle with precise enumeration, often exhibiting ``numerosity blindness'' or hallucination when counts exceed small subitizing ranges ($N < 5$)~\cite{clipcount_paiss_iccv2023,sugarcrepe_hsieh_nips2023,vlmblind_rahmanzadehgervi_acvv2024,hallusionbench_guan_cvpr2024,paligemma_beyer_arxiv2025,vlmcount20_guo_arxiv2025}. 
Our work extends this line of inquiry by moving beyond performance benchmarking to identifying the precise architectural stage: recognition, numerosity, or articulation, where this capability collapses.

\subsection{Systematic Generalization and Extrapolation}
Systematic generalization, the ability to apply learned rules to inputs outside the training distribution, is a known weakness of neural networks. 
In Natural Language Processing (NLP), this is often framed as ``length generalization'': models trained on short arithmetic or logical sequences fail to process longer sequences at test time~\cite{systemgeneralization_lake_icml2018,composedecompose_hupkes_jair2020,traintestlen_press_iclr2022,explorelen_anil_nips2022}. 
Theoretical analyses suggest that Transformers tend to overfit to positional correlations rather than learning the recursive underlying algorithms~\cite{devilisdetail_csordas_emnlp2021,nnchomsky_deletang_iclr2023,poslen_kazemnejad_nips2023,transformernotstructural_zhang_tmlr2024}. 
In the multimodal domain, this challenge manifests as \textit{Visual Extrapolation}: the inability to count quantities unseen during training. 
Unlike prior works that attempt to solve this via data scaling or prompting strategies~\cite{cot_wei_nips2022,multimodalcot_zhang_arxiv2023}, we focus on the mechanistic failure of cross-modal grounding, demonstrating that the failure to extrapolate is not a loss of knowledge but a failure of alignment between modalities.

\subsection{Probing Multimodal Representations}
Understanding the internal state of ``black box'' models is critical for diagnosing reasoning failures. 
Linear probing has emerged as a standard technique to extract latent information from frozen representations, revealing that models often encode features (e.g., syntax, semantics) that are not explicitly supervised~\cite{linearprob_alain_iclr2017,embprobe_adi_iclr2017,bertprobe_tenney_acl2019,controlprobe_hewitt_emnlp2019}. 
In the context of VLMs, probing has been used to assess spatial awareness, attribute binding, and object existence~\cite{windoprobe_thrush_cvpr2022,vlmbog_yuksekgonul_iclr2023,hallucinationvlm_li_emnlp2023}. 
Closely related to our work, \cite{nlpnum_wallace_emnlp2019,neuronnumprob_gurnee_tmlr2023} investigated the ``number sense'' of language models, finding that number neurons exist but are often polysemantic. 
Our contribution diverges by applying probing specifically to the \textit{magnitude extrapolation} regime. 
We provide the first evidence of a ``Fractured Magnitude'' phenomenon, where the visual encoding of number remains robust and linear even when the symbolic mapping of that number collapses entirely.

\section{Limitations} 
\label{sec:limitations}
While our work offers robust mechanistic evidence for the \textit{Fractured Magnitude Hypothesis}, we acknowledge several limitations in our experimental scope:

\paragraph{Assumption of Perceptual Success.} Our evaluation framework focuses strictly on the algorithmic capacity to enumerate quantities, factoring out the confounding variable of object recognition. In wild scenarios, counting failures often stem from detection failures (e.g., occlusion or camouflage). We do not penalize the model for failing to recognize the target object class; however, our Stage 1 probing results confirm that for the chosen domain, the visual signal is present, justifying this isolation of the reasoning bottleneck.

\paragraph{Complexity of Visual Stimuli.} We utilize Go game boards as a controlled environment to ensure precise ground truth and strictly enforce identical objects. Real-world visual counting is significantly more complex, requiring semantic filtering (e.g., ``count the red apples but ignore the green ones'') and handling high intra-class variance. While our simplified setting yields a necessary upper bound on counting capability, it may not fully capture the noise and ambiguity of natural image datasets.

\paragraph{Model Availability and Scope.} Our real-world validation is primarily conducted on Qwen3-VL-32B. While this was necessary because other open-weights models (e.g., LLaVA, previous Qwen iterations) lacked sufficient baseline counting capabilities to meaningfully test extrapolation, our conclusions strictly apply to this architecture. However, given that Qwen3-VL represents the current state-of-the-art in open-source VLMs, we believe these findings are indicative of broader architectural limitations in the Transformer paradigm.

\paragraph{Task Specificity vs. Generalization.} Our analysis focuses on counting black stones, a specific feature that may appear with varying frequency in pretraining corpora. It is possible that VLMs exhibit better performance on highly frequent objects (e.g., ``people'' or ``cars'') due to memorized density-label correlations. However, our objective is to measure the model's \textit{expressive power} and algorithmic generalization rather than its retrieval capacity. The failure to extrapolate on stones despite having a clear visual signal suggests a fundamental architectural deficit rather than a lack of domain-specific training data.

\section{Use of AI Tools}
This work utilized Gemini for linguistic refinement, improving text clarity, formatting \LaTeX\ code. 
AI models were strictly limited to an assistive role and were not involved in research ideation, experimental design, theoretical derivation, or the interpretation of results.

\section{Supplementary}
\label{app:sup}
In this section, we provide details about our supplementary experiment results and our experiment settings.

\subsection{Experiment Results}
\label{app:sup:results}
We provide additional experiment results below.

\subsubsection{Causal Analysis}
\label{app:sup:results:causal}
While linear probes confirm that magnitude information ($N_H$) is linearly separable within the Vision Encoder, probing is correlational; it does not guarantee that the model actively utilizes these specific features for its prediction. 
We aim to establish a rigorous \textit{causal link} between the probed visual representations and the generated count ($N_P$). 
If the features detected by our probe are indeed the drivers of counting, surgically suppressing them should result in a deterministic and predictable drop in the model's output.

\paragraph{Experiment Design.}
We perform an activation steering (intervention) experiment on the frozen Vision Encoder outputs using the In-Distribution dataset ($N_G \in [0, 49]$), where the model's baseline counting is $100\%$ accurate.
For a given input image, we utilize the probe to identify the patch embeddings corresponding to $k$ specific black stones (where $k$ ranges from $1$ to $5$).
We mask the corresponding tokens for the $k$ black stone in attention calculation.
If this token is indeed responsible for the black stone counting, the counting number should be $k$ smaller than the original number.
Thus, we define the \textit{Steering Accuracy} as the percentage of cases where the model's new prediction $N_P'$ satisfies the subtraction logic: $N_P' = N_G - k$.

\begin{figure}[!ht]
    \centering
    \scalefont{0.8}
    \begin{tikzpicture}
    \begin{axis}[
        ybar,
        width=8cm, height=4.5cm, %
        xlabel={Intervention Number ($k$)},
        ylabel={Accuracy for $N_P'=N_G-k$ (\%)},
        ymin=0, ymax=120, %
        symbolic x coords={1, 2, 3, 4, 5},
        xtick=data,
        every node near coord/.append style={font=\tiny, rotate=90, anchor=west}, 
        xticklabel style={font=\scriptsize, rotate=90},
        ylabel near ticks,
        xlabel near ticks,
        bar width=0.5cm, %
        enlarge x limits=0.2,
        ymajorgrids=true,
        grid style=dashed,
    ]

    \addplot+[
        fill=ETHBlue, 
        draw=none,
        opacity=0.8
    ] coordinates {
(1, 99.73)
(2, 99.38)
(3, 96.78)
(4, 86.89)
(5, 73.25)
    };

    \end{axis}
    \end{tikzpicture}
    \caption{\textbf{Black-stone patch embeddings causally drive counting via attention masking.} 
    On the In-Distribution set (baseline counting accuracy: $100\%$), we randomly sample k specific black stones in each image $(k \in [1, 5])$, and mask the corresponding tokens in the attention computation.
    We report \textbf{Steering Accuracy}, defined as the percentage of examples where the intervened prediction matches the new groundtruth $N_G' = N_G - k$.
    The model achieves near-perfect Steering Accuracy for $k \in [1, 3]$, shows the first noticeable degradation at $k = 4$, and drops to $73.25\%$ at $k = 5$, consistent with attention distribution shifts when masking many tokens.}
    
    \label{fig:syn:causal}
\end{figure}
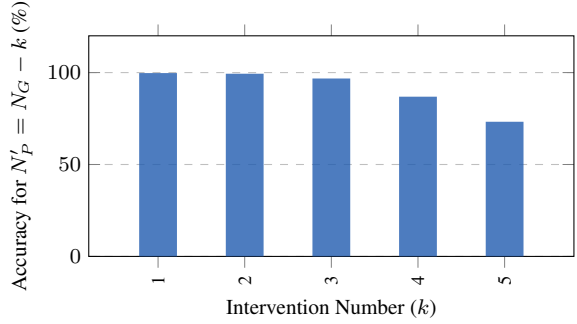

\paragraph{Experiment Results.}
As shown in \cref{fig:syn:causal}, the results are definitive. 
For every intervention size $k \in [1, 4]$, the model achieves almost $100\%$ Steering Accuracy. 
When $k=5$, the Steering Accuracy is only $77.2\%$. We presume this is caused by the attention distribution shift when masking that many tokens.
For example, replacing the visual representation of $k=3$ black stones with background noise causes the model to decrement its count by exactly $N_P' = N_G - 3$ in every single instance. 
This indicates a precise, one-to-one functional dependency between the presence of these specific feature vectors and the arithmetic sum computed by the model.

\paragraph{Takeaway.}
This provides causal evidence that the model is not relying on spurious correlations or global texture statistics. 
The \textit{hidden number} features identified by our probes are the true mechanistic inputs to the counting algorithm. 
Consequently, the downstream extrapolation failure cannot be attributed to the model ignoring the visual signal or attending to the wrong features; the model correctly ingests the ``bricks'' of the count, confirming the failure occurs strictly in the decoding logic.

\subsubsection{Layer-Wise Hidden Number Probing}
\label{app:sup:results:syn_probe}

While the \textit{Comparative Counting} task (\cref{sec:syn:stage2}) demonstrates that the model can utilize magnitude information for binary decision-making, it treats the internal representations as a black box. 
To rigorously verify \textit{Hypothesis B}, we need to determine if the ``Hidden Number'' ($N_H$) is preserved as a linearly separable feature throughout the network's depth, or if the signal-to-noise ratio degrades as it propagates through the deep Transformer layers of the language decoder.

\paragraph{Experiment Design.}
We extend the linear probing methodology described in \cref{sec:framework:probe} to a layer-wise analysis. 
We train a separate linear classifier $f_{\text{probe}}^{(l)}$ on the output hidden states of every layer $l$ in the architecture:
\begin{itemize}
    \item \textbf{Vision Encoder:} Layers 1--2 (Final output used for Stage 1 analysis).
    \item \textbf{Language Decoder:} Layers 1--2 (The reasoning backbone).
\end{itemize}
The probes are trained on the In-Distribution (ID) dataset ($N \le 49$) and evaluated on the Visual Extrapolation (VE) dataset ($N \in [50, 99]$). 
High classification accuracy at layer $l$ indicates that the magnitude information remains distinct and retrievable at that depth.

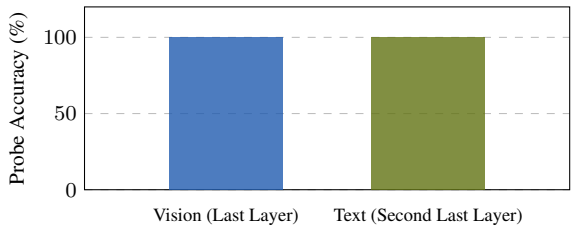
\begin{figure}[!ht]
    \centering
    \scalefont{0.8}
    \begin{tikzpicture}
    \begin{axis}[
        ybar,
        width=8cm, height=4cm,
        ylabel={Probe Accuracy (\%)},
        ymin=0, ymax=120,
        symbolic x coords={Vision (Last Layer), Text (Second Last Layer)},
        xtick={Vision (Last Layer), Text (Second Last Layer)},
        nodes near coords align={vertical},
        every node near coord/.append style={font=\tiny,}, 
        xticklabel style={font=\scriptsize},
        ylabel near ticks,
        tick style={draw=none},
        bar width=1.5cm,
        enlarge x limits=.7,
        ymajorgrids=true,
        grid style=dashed,
    ]

    \addplot+[
        fill=ETHBlue,
        draw=none,
        opacity=0.8,
        bar shift=0pt,
    ] coordinates {
        (Vision (Last Layer), 100)
    };
    
    \addplot+[
        fill=ETHGreen,
        draw=none,
        opacity=0.8,
        bar shift=0pt,
    ] coordinates {
        (Text (Second Last Layer), 100)
    };
    \end{axis}
    \end{tikzpicture}
    \vspace{-.2cm}
    \caption{\textbf{Latent quantity representations are perfectly linearly separable.} 
    Linear probes trained on both the vision and language modules achieve 100\% accuracy. 
    This confirms that distinct magnitude information persists across modalities, ruling out representation collapse as the failure mechanism.}
    \label{fig:syn:quantity:probe}
    \vspace{-.3cm}
\end{figure}

\paragraph{Results.}
As illustrated in \cref{fig:syn:quantity:probe}, the probes achieve $100\%$ accuracy across all layers of both the Vision Encoder and the Language Decoder. 
This result is highly significant: it confirms that the representation of quantity does not ``drift'' or become entangled as it passes through the cross-modal attention layers. 
For every number $N \in [0, 99]$, the model maintains a precise, linearly separable manifold in the high-dimensional latent space.

\paragraph{Takeaway.}
These findings provide structural confirmation for the rejection of \textit{Hypothesis B}. 
The failure to articulate the number ``75'' cannot be attributed to a loss of information or reasoning capacity, as the exact integer value is perfectly encoded in the activation space of the very last layer before the final vocabulary projection.

\subsection{Experiment Settings}
\label{app:sup:settings}
We introduce details about the synthetic VLM training and the prompt templates below.

\subsubsection{Synthetic VLM Training}
\label{app:sup:settings:training}

To ensure reproducibility, we provide the specific architectural configurations and training hyperparameters used for synthetic experiments. The setup applies to both the generative counting task and the comparative (True/False) reasoning task.

\paragraph{Model Architecture.}
We utilize a lightweight, custom Vision-Language Model designed to isolate mechanistic behaviors.
\begin{itemize}
    \item \textbf{Vision Encoder:} A 2-layer Transformer with a patch size of $14 \times 14$.
    \item \textbf{Language Decoder:} A 2-layer Causal Transformer.
    \item \textbf{Dimensions:} Both modules use a hidden embedding dimension of $d_{\text{model}} = 32$ with $H=2$ attention heads. 
    \item \textbf{Positional Embeddings (M-RoPE):} Following the setting of Qwen3-VL-Instruct~\citep{qwen3vl_bai_arxiv2025}, we employ Multimodal Rotary Positional Embeddings (M-RoPE)~\citep{mrope_huang_arxiv2025} to encode spatiotemporal information. The rotary dimensions are allocated as sections $[4, 6, 6]$, corresponding to the proportions of the head dimension ($d_{\text{head}} = 16$) assigned to time ($t$), height ($h$), and width ($w$) respectively.
\end{itemize}

\paragraph{Tokenization.}
We employ a custom tokenizer to handle numerical reasoning explicitly:
\begin{itemize}
    \item \textbf{Whitespace Separation:} Input text is tokenized based on whitespace.
    \item \textbf{Digit Splitting:} Crucially, we enable \texttt{individual\_digits=True}, ensuring that multi-digit numbers are tokenized as sequences of discrete digits (e.g., ``100'' $\rightarrow$ [``1'', ``0'', ``0'']). This prevents the model from treating large numbers as unique atomic vocabulary items, forcing it to learn place-value counting rules.
\end{itemize}

\paragraph{Training Protocol.}
We adopt a two-stage curriculum with a total budget of 20 epochs per stage.
\begin{itemize}
    \item \textbf{Data Scale:} For both the text-only pretraining and the visual counting training phase, we generate 8,192 training examples for each target number class.
    \item \textbf{Optimization:} We train using a batch size of 2,048 with an initial learning rate of $lr = 1e^{-3}$. For the finetuning, the vision encoder has $lr = 1e^{-3}$ and language decoder has  $lr = 1e^{-5}$.
    \item \textbf{Curriculum:} 
    \begin{itemize}
        \item \textit{Phase 1 (Text Priors):} Pretraining on text-only numerical sequences to establish counting algorithms.
        \item \textit{Phase 2 (Visual Alignment):} %
        Training on visual counting data (mixed with Phase 1 textual data to prevent forgetting) to align visual features with the pretrained numerical representations.
    \end{itemize}
\end{itemize}

\subsubsection{Prompt}
\label{app:sup:settings:prompt}
We provide prompt template used in both synthetic and real VLM below.

\begin{figure}[!ht]
\begin{tcolorbox}[colback=violet!10, colframe=violet!70!black, title={\large Prompt Template for Text Counting with Synthetic VLM}]

\begin{tcolorbox}[colback=gray!10, colframe=gray!80]
\textbf{Input String:} 

[a a b c a c b c c b b b c b a a b c b b b b b a b a a c b b a b b c b b a b b b a a c b b c c c b b b a b c b c b c a c c a b c c c b a a b b b a c b c b b b b a b c a c a b a b c b a b c c b b a c b a c b a b a b c c b a c a b b b a a c b c b c b b b c c b a a b b c c c b c b a a a c a c b a a b b a b c b a b a b a b a c b b b a c c c c a c b b b c c a a b b a a b a c b c a a c c b b a b b c a c c b b b b c b c b a a c b a c c b b b b c b b a c b a a a a b a b a c a c b b a a a b a b a a a b a b a a a a a c c a b a a b a b b c b b b a b b a a c c b c a a a b c b a a a c a b a a a a b a b a b b a c c c a b c c a a c b b c b a b a a c b c a a b a a c b c b a c c c c b a b a c b c b c b b a c a b a b a b c b a b b]
\end{tcolorbox}

\begin{tcolorbox}[colback=gray!10, colframe=gray!80]
\setstretch{1.1}

\textbf{Question:}
How many c letters are there in the given string ?

\end{tcolorbox}

\begin{tcolorbox}[colback=gray!10, colframe=gray!80]
\textbf{Answer:} 
96
\end{tcolorbox}

\end{tcolorbox}
\caption{Prompt template for text counting with synthetic VLM.}
\label{fig:prompt_template:syn:text}
\end{figure}

\begin{figure}[!ht]
\begin{tcolorbox}[colback=violet!10, colframe=violet!70!black, title={\large Prompt Template for Vision Counting with Synthetic VLM}]

\begin{tcolorbox}[colback=gray!10, colframe=gray!80]
\textbf{Input:} 
\begin{center}
    \includegraphics[width=0.38\textwidth]{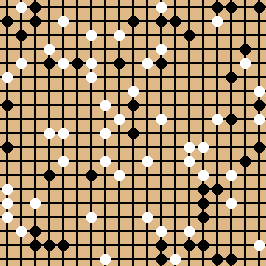}
\end{center}
\end{tcolorbox}

\begin{tcolorbox}[colback=gray!10, colframe=gray!80]
\setstretch{1.1}

\textbf{Question:}
How many black stones are there on the chessboard ?

\end{tcolorbox}

\begin{tcolorbox}[colback=gray!10, colframe=gray!80]
\textbf{Answer:} 
41
\end{tcolorbox}

\end{tcolorbox}
\caption{Prompt template for vision counting with synthetic VLM.}
\label{fig:prompt_template:syn:vision}
\end{figure}

\begin{figure}[!ht]
\begin{tcolorbox}[colback=violet!10, colframe=violet!70!black, title={\large Prompt Template for Text Counting Comparison with Synthetic VLM}]

\begin{tcolorbox}[colback=gray!10, colframe=gray!80]
\textbf{Input:} 

[a a b c a c b c c b b b c b a a b c b b b b b a b a a c b b a b b c b b a b b b a a c b b c c c b b b a b c b c b c a c c a b c c c b a a b b b a c b c b b b b a b c a c a b a b c b a b c c b b a c b a c b a b a b c c b a c a b b b a a c b c b c b b b c c b a a b b c c c b c b a a a c a c b a a b b a b c b a b a b a b a c b b b a c c c c a c b b b c c a a b b a a b a c b c a a c c b b a b b c a c c b b b b c b c b a a c b a c c b b b b c b b a c b a a a a b a b a c a c b b a a a b a b a a a b a b a a a a a c c a b a a b a b b c b b b a b b a a c c b c a a a b c b a a a c a b a a a a b a b a b b a c c c a b c c a a c b b c b a b a a c b c a a b a a c b c b a c c c c b a b a c b c b c b b a c a b a b a b c b a b b]

[c b a b b b a c c b b b c b b c b c c a a a c c c a a b a a a a a b b a c b b b b b a b b a b b b b b a c a b b a b b c b c c b a b b b c b c c b c b b b b c c b a b c a b a b a b c c a a c b a c a b c a b a c a a a a b c b a a c a a a c a a c c a b c a c a c c a b a a b a a b b b b a a a a b b b b a c a a a a b b a c b b c a c c c b c c a c b b c c b a b c b c c a c b a c b a a b a a b b b c b a a a b b a b b b c c c b c c b b b a b b b a b c b b b a c b c b a b b b b c c a a b b c b a c c a a c b b a a b c a c b b a a b b b c a c b a b a c a b a b a c b b a b a b b b c b c a b a c c b b a b b a c c b c a b b b c b c a c c b c b c b c b a b b a b a a a a c a b c a b c b a a a c b a a b a a c a c b c c a a a b b]

\end{tcolorbox}

\begin{tcolorbox}[colback=gray!10, colframe=gray!80]
\setstretch{1.1}

\textbf{Question:}
Are the number of c letters in both input strings the same ?

\end{tcolorbox}

\begin{tcolorbox}[colback=gray!10, colframe=gray!80]

\textbf{Answer:} True
\end{tcolorbox}

\end{tcolorbox}
\caption{Prompt template for text counting comparison with synthetic VLM.}
\label{fig:prompt_template:syn:text_compare}
\end{figure}

\begin{figure}[!ht]
\begin{tcolorbox}[colback=violet!10, colframe=violet!70!black, title={\large Prompt Template for Vision Counting Comparison with Synthetic VLM}]

\begin{tcolorbox}[colback=gray!10, colframe=gray!80]
\textbf{Input:} 

\begin{center}
    \includegraphics[width=0.38\textwidth]{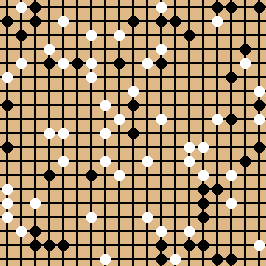}
\end{center}

[c b a b b b a c c b b b c b b c b c c a a a c c c a a b a a a a a b b a c b b b b b a b b a b b b b b a c a b b a b b c b c c b a b b b c b c c b c b b b b c c b a b c a b a b a b c c a a c b a c a b c a b a c a a a a b c b a a c a a a c a a c c a b c a c a c c a b a a b a a b b b b a a a a b b b b a c a a a a b b a c b b c a c c c b c c a c b b c c b a b c b c c a c b a c b a a b a a b b b c b a a a b b a b b b c c c b c c b b b a b b b a b c b b b a c b c b a b b b b c c a a b b c b a c c a a c b b a a b c a c b b a a b b b c a c b a b a c a b a b a c b b a b a b b b c b c a b a c c b b a b b a c c b c a b b b c b c a c c b c b c b c b a b b a b a a a a c a b c a b c b a a a c b a a b a a c a c b c c a a a b b]

\end{tcolorbox}

\begin{tcolorbox}[colback=gray!10, colframe=gray!80]
\setstretch{1.1}

\textbf{Question:}
Is the number of c letters in the given string the same as the number of black stones on the chessboard ?

\end{tcolorbox}

\begin{tcolorbox}[colback=gray!10, colframe=gray!80]

\textbf{Answer:} False
\end{tcolorbox}

\end{tcolorbox}
\caption{Prompt template for vision counting comparison with synthetic VLM.}
\label{fig:prompt_template:syn:vision_compare}
\end{figure}

\begin{figure}[!ht]
\begin{tcolorbox}[colback=violet!10, colframe=violet!70!black, title={\large Prompt Template for Text Counting with Real VLM}]

\begin{tcolorbox}[colback=gray!10, colframe=gray!80]
\textbf{System Prompt:} 
You are a precise counting engine. 1. Verification: Before answering, output a compact data check (e.g., list of indices for text, or row-by-row counts for images). Avoid conversational filler. 2. Format: Verification: $<$Dense Data$>$ $\rightarrow$ FINAL\_ANSWER: $<$Result$>$.
\end{tcolorbox}

\begin{tcolorbox}[colback=gray!10, colframe=gray!80]
\textbf{Input:} 

[`c', `b', `a', `a', `b', `a', `a', `a', `b', `c', `c', `c', `c', `a', `c', `c', `c', `b', `b', `a', `c', `c', `a', `a', `c', `a', `a', `a', `c', `c', `a', `b', `c', `a', `c', `b']

\end{tcolorbox}

\begin{tcolorbox}[colback=gray!10, colframe=gray!80]
\setstretch{1.1}

\textbf{Question:} 
String length: 36. Count the occurrences of letter 'c'.

\end{tcolorbox}

\begin{tcolorbox}[colback=gray!10, colframe=gray!80]

\textbf{Answer:} 
15

\end{tcolorbox}

\end{tcolorbox}
\caption{Prompt template for text counting with real VLM. }
\label{fig:prompt_template:real:text}
\end{figure}

\begin{figure}[!ht]
\begin{tcolorbox}[colback=violet!10, colframe=violet!70!black, title={\large Prompt Template for Vision Counting with Real VLM}]

\begin{tcolorbox}[colback=gray!10, colframe=gray!80]
\textbf{System Prompt:} 
You are a precise counting engine. 1. Verification: Before answering, output a compact data check (e.g., list of indices for text, or row-by-row counts for images). Avoid conversational filler. 2. Format: Verification: $<$Dense Data$>$ $\rightarrow$ FINAL\_ANSWER: $<$Result$>$.
\end{tcolorbox}

\begin{tcolorbox}[colback=gray!10, colframe=gray!80]
\textbf{Input:} 
\begin{center}
    \includegraphics[width=0.12\textwidth]{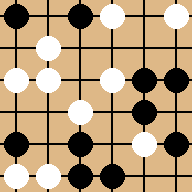}
\end{center}

\end{tcolorbox}

\begin{tcolorbox}[colback=gray!10, colframe=gray!80]
\setstretch{1.1}

\textbf{Question:} 
Analyze the 6x6 Go board. Count the black stones.

\end{tcolorbox}

\begin{tcolorbox}[colback=gray!10, colframe=gray!80]

\textbf{Answer:} 
10

\end{tcolorbox}

\end{tcolorbox}
\caption{Prompt template for vision counting with real VLM. }
\label{fig:prompt_template:real:vision}
\end{figure}

\end{document}